\documentstyle[prl,aps,epsf,multicol]{revtex}
\begin{document}
\draft

\newcommand{\sigmatwo}{\sigma_e}
\newcommand{\dcross}{\mbox{$d\!\!^-$}}
\newcommand{\intk}{\int\dcross^2 \!k\dcross\! \omega\,}
\newcommand{\intKthree}{\int\dcross^3\!K\dcross\!\Omega\,}
\newcommand{\intK}{\int\dcross^2\!K\dcross\!\Omega\,}
\newcommand{\psibar}{\bar{\psi}}

\def\i{\imath\,}
\def\ih{\frac{\imath}{2}\,}
\def\undertext#1{\vtop{\hbox{#1}\kern 1pt \hrule}}
\def\ra{\rightarrow}
\def\lfa{\leftarrow}
\def\ua{\uparrow}
\def\da{\downarrow}
\def\Ra{\Rightarrow}
\def\lra{\longrightarrow}
\def\ler{\leftrightarrow}
\def\lrb#1{\left(#1\right)}
\def\O#1{O\left(#1\right)}
\def\EV#1{\left\langle#1\right\rangle}
\def\tr{\hbox{tr}\,}
\def\trb#1{\tr\lrb{#1}}
\def\dd#1{\frac{d}{d#1}}
\def\dbyd#1#2{\frac{d#1}{d#2}}
\def\pp#1{\frac{\partial}{\partial#1}}
\def\pbyp#1#2{\frac{\partial#1}{\partial#2}} 
\def\pd#1{\partial_{#1}}
\def\br{\\ \nonumber & &}
\def\brr{\right. \\ \nonumber & &\left.}
\def\inv#1{\frac{1}{#1}}
\def\be{\begin{equation}}
\def\ee{\end{equation}}
\def\bea{\begin{eqnarray}}
\def\eea{\end{eqnarray}}
\def\ct#1{\cite{#1}}
\def\rf#1{(\ref{#1})}
\def\EXP#1{\exp\left(#1\right)} 
\def\INT#1#2{\int_{#1}^{#2}} 
\def\LHS{left-hand side }
\def\RHS{right-hand side }
\def\COM#1#2{\left\lbrack #1\,,\,#2\right\rbrack}
\def\AC#1#2{\left\lbrace #1\,,\,#2\right\rbrace}

\title{Screening and Dissipation at the Superconductor-Insulator 
Transition \\ Induced by a Metallic Ground Plane}
\author{Ashvin Vishwanath$^{(1)}$ \and Joel E. Moore$^{(2)}$ and T. Senthil$^{(1)}$ } 

\address{$^{(1)}$Department of Physics, Massachusetts Institute of Technology, 
77 Massachusetts Ave., Cambridge MA 02139. \\${(2)}$Department of
Physics, University of California, Berkeley, CA 94720; Materials
Sciences Division, Lawrence Berkeley National Laboratory, Berkeley, CA
94720}

\date{\today}
\maketitle

\begin{abstract}
We study localization phenomena in two dimensional systems of charged
particles in the presence of a metallic ground plane with a
particular focus on the superconductor-insulator transition. The
ground plane introduces a screening of the long-range Coulomb
interaction, and provides a source of dissipation due to the gapless
diffusive electrons.  The interplay of these two effects leads to
interesting physical phenomena which are analysed in detail in this
paper.  We argue that the generic superconductor-insulator transition
of charged particles in the presence of the ground plane may be
controlled by a {\it fixed line} with variable critical
exponents. This is illustrated by an explicit calculation in an
appropriate $\epsilon$ expansion. In contrast, the universal
properties of the superconductor-Mott insulator transition in the
clean limit at commensurate densities are shown to be unmodified by
either the metal or the long-range Coulomb interaction. A similar
fixed line can arise in the presence of a metallic ground plane for
quantum Hall plateau transitions.
Implications for experiments on Josephson-junction arrays and quantum
Hall systems are described.
\end{abstract}

\vspace{0.15cm}


\begin{multicols}{2}
\narrowtext
\section{Introduction}

Despite many decades of effort, the properties of matter in the
vicinity of various transitions from conducting to insulating phases
remains poorly understod. The best known example is the transition
from a metal to an insulator. During the last few years, a number of
such phenomena have been subjected to serious experimental study,
particularly in low dimensional systems.  These include transitions
from a superconductor to an insulator \cite{Goldman}, various
transitions in two dimensional systems showing the quantum Hall effect
\cite{DasSharma}, and an apparent transition from an as yet
unidentified conducting phase to an insulator in two dimensional
Silicon MOSFET's \cite{Kravchenko}. This increase in experimental
activity has raised questions that require an improved theory of such
localization phenomena.

In this paper, we study localization phenomena in two dimensional
systems in the presence of a proximate ground plane formed by a
diffusive two dimensional electron gas. Our focus will be the
superconductor-insulator transition, though many of our results
generalize to other transitions as well.  Recent experiments have
studied the effect of such a metallic plane on the quantum transition
to superconductivity in Josephson junction arrays \cite{Rimberg} and
in homogenously disordered thin films in a magnetic field \cite{Mason}
. In these experiments, the actual tunneling of charged excitations
between the superconductor and the metallic plane is very weak, and
may be ignored.  However, the charge carriers in the two systems are 
coupled together by the Coulomb interaction. Such a metallic plane has
several effects. First, one might expect some screening of the
long-ranged Coulomb interaction between the charges in the system
undergoing the localization transition.  Second, the coupling to the
gapless excitations in the metal provides a source of ``dissipation''.
The interplay of these two effects leads to interesting differences in
the localization phenomena from that in the absence of the metallic
plane.  In particular, we argue that the zero temperature localization
transition may be controlled by a {\em fixed line} with variable
critical exponents.

The interest in studying the role of a metallic plane stems from
several different directions.  A particularly difficult issue for
localization theory is the role of the long-range Coulomb interaction
and its possible screening.  For the integer quantum Hall transitions,
for instance, while short-range interactions have been argued to be
irrelevant at the non-interacting fixed point \cite{dhlee}, the nature
of the criticality with Coulomb interaction is very poorly understood.
For metal-insulator or superconductor-insulator transitions, the
Coulomb interaction is screened in the conducting phase but retains
its long-range character in the insulating phase. This evolution in
the screening properties of the system as it moves through the
transition contributes to the difficulty in developing a theoretical
understanding. Introducing a metallic plane enables exerting some
control on the screening and may enable separating out phenomena that
are specific to the long-range Coulomb interaction.

A different motivation comes from theoretical attempts to understand
experiments on the destruction of superconductivity at zero
temperature due to quantum effects. In situations where the transition
is to an insulator, and at asymptotically low temperatures, a
``boson-only'' model that describes the localization of Cooper pairs
while ignoring the fermionic degrees of freedom is expected to
describe the universal critical properties \cite{Fisher}. However, in
a number of experiments, the situation is different
\cite{Goldman,Mason}. The non-superconducting phase may be a weakly
localized metal, or in some cases appear to be truly metallic (with no
trace of localization effects) in the temperature range currently
probed in experiments.  In these cases, a ``boson-only'' model is
presumably inadequate, and it is necesary to include the underlying
electrons. However a full treatment of the problem of Cooper pairs
coupled to gapless electrons is prohibitively difficult---it therefore
helps to study simpler situations where some if not all the effects
are treated reliably. Such a simplification is provided by considering
``boson-only'' models coupled to gapless metallic electrons through
density-density Coulomb interactions, in which the decay of the boson
into pairs of electrons is ignored.  Precisely this situation is
realized in experiments on Josephson junction arrays where a proximate
metallic plane is introduced \cite{Rimberg}.  Remarkably, as we show
in this paper, such a coupling between the Cooper pairs and the
metallic electrons has a profound effect on the properties of the
transition.

The nature of the superconducting phase and its transition to the
insulator in the presence of a metallic plane has been addressed
before in the literature.  We believe that these previous treatments
miss several crucial aspects of the physics.  Ref. \cite{Wgblst1}
considered a model of a Josephson junction array with a short-ranged
capacitance matrix that essentially amounts to assuming a {\em
logarithmic} Coulomb interaction energy between the charges rather
than the $1/r$ interaction that obtains for long distances and controls the universal properties of the transition. This difference
profoundly modifies the physics--- hence the results of
Ref. \cite{Wgblst1} are not capable of describing the experiments of
interest\cite{Rimberg,Mason}.  There are also a number of papers on
``local dissipative'' models, where the
dissipation is claimed to arise from spatially localized excitations.  In a short section below, we
discuss the relationship between these models and our work. First in our work unlike in Ref. \cite{Wgblst2}, the
physical origin of the dissipation is clear - it is due to the gapless diffusive electrons in the metal. 
Furthermore, as we discuss in detail,  the fixed line obtained here has an entirely
{\em different} origin from the fixed line claimed to exist in the model studied in 
Ref. \cite{Wgblst2} (see also \cite{Volker}).  Deep inside the
superconducting phase, the dynamics of the phase of the
superconducting order parameter in the presence of the coupling to the
metal was considered by Gaitonde \cite{Gaitonde}, who argued that the
plasmon was overdamped at long wavelengths.  We derive the correct
phase dynamics and show that in fact the plasmon mode survives at long
wavelengths.

For most of this paper, we focus on the superconductor-insulator
transition of repulsively interacting bosons in the presence of
density-density coupling to a diffusive two dimensional electron
gas. We begin by first considering the effects of the metallic plane
on the superconducting phase. We show that the plasmon mode of the
superconductor survives essentially unmodified. This is contrary to
naive expectations based on the assumption that a short-ranged
screened interaction between the bosons correctly describes {\em all}
aspects of the physics with the metal present.  We find that, because
of their slow diffusive motion, the electrons in the metal are unable
to screen out the faster charge fluctuations associated with the
plasmon. This analysis sets the stage for a discussion of the general
properties of the transition to the insulator. With the metallic plane
present, we give a general argument that points to the existence of a
{\em fixed line} controlling the transition. The fixed line is
parametrized by the value of the conductivity of the metal. Universal
critical properties will then be determined by the metal conductivity.

To substantiate our general arguments we present a detailed analysis
of a specific model which is appropriate to describe the experiments
in Ref. \cite{Rimberg}.  We consider the superconductor-insulator
transition at commensurate density for bosons on a square lattice in
the presence of the metal.  We first provide crude estimates of the
phase boundary and show that the presence of the metal stabilizes the
superconducting phase relative to the insulator. We derive an
effective action for the Cooper pair degrees of freedom by integrating
out the diffusive metallic electrons which is then used to discuss the
critical phenomena at the transition. In the clean system, we argue
that neither the metal nor the long-ranged part of the Coulomb
interaction affects the critical properties so that the transition is
in the universality class of the $D = 2 + 1$ $XY$ model. This result
is strongly suggested by two separate calculations---a large-$N$
generalization of our effective model, and an $\epsilon$ expansion of
an appropriate generalization to dimensions other than $2$. However
moving away from this idealized limit is expected to alter the
critical properties and produce the fixed line expected from the
general arguments advanced earlier. We illustrate this by including
disorder which nevertheless preserves the particle/hole symmetry of
the commensurate problem. A double $\epsilon$ expansion \cite{Cardy}
along the lines of Ref. \cite{Herbut} is then performed, including
both the Coulomb interaction and the effect of the metal, which indeed
finds the expected fixed line.

The generality of our arguments for the existence of a fixed line in
the presence of the metal is then illustrated in the context of other
concrete yet tractable models of localization transitions.
Specifically, we show that a model for the quantum Hall transition
considered in Ref. \cite{yesubir} should also possesses this
feature. Indeed we expect that the generic quantum Hall transition
with long-range Coulomb interactions could, in the presence of a
proximate metal, also be controlled by a fixed line with variable
critical exponents.

The implications of our results for experiments on Josephson-junction
arrays, superconducting films, and quantum Hall transitions are considered in Section
\ref{exp_imp}.
An important qualitative lesson is that the presence of a proximate
metallic plane {\em does not} by itself guarantee that the
universality class is that of the short-ranged problem. Under what
circumstances then is the latter realized? For the generic
superconductor-insulator or quantum Hall transition, our analysis
indicates that when the fixed line scenario is realized, and the
conductivity of the metal is very large (compared to $e^2/h$) there
will be a region near the transition where the universal behaviour is
controlled by the short-range fixed point. Very close to the critical
point however there will be a crossover to behaviour controlled by the
appropriate point on the fixed line.  The presence of a fixed line
directly impacts experiments seeking to measure universal transport at
these two dimensional localization transitions.  Indeed, a
temperature-independent conductivity will obtain right at the
transition point, with a value that will vary along the fixed line and
hence may appear to be non-universal. However, this conductivity will
be universally related to the critical exponents (which too will vary
along the fixed line). Experimental realization of the fixed line may be easier 
to observe for the quantum hall transition than for the superconductor-insulator 
transition as the universal scaling regime appears to be more easily accesible in the former. 

It is interesting to speculate on the implications of our results for
the general superconductor-metal transition. Our analysis shows that a
simple Coulomb coupling between the order parameter (boson) degrees of
freedom and gapless diffusive electrons already leads to profound
modifications of the universal properties from that of a boson-only
model. To correctly describe the superconductor-metal transition , it
is necessary to include processes where the Cooper pairs can decay
into a pair of electrons and vice versa into the model. If any portion
of the fixed line survives this inclusion, then the conductivity at
the transition will again appear to be non-universal and will be
determined by the value of the marginal coupling at the point on the
fixed line that controls the transition. We note that experiments on
two-dimensional superconductors undergoing a transition to a weakly
localized metal do see a temperature-independent but apparently
non-universal conductivity at the transition.
 
\section{General Arguments}
\begin{figure}
\epsfxsize=2.8in
\centerline{\epsffile{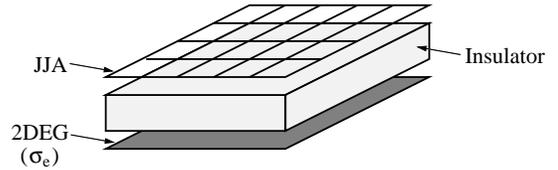}}
\vspace{0.1in}
\caption{The setup considered in this paper, based on the experiments of 
Rimberg et al.[4]  A Josephson junction array (JJA) is separated from a
two dimensional electron gas (with conductivity $\sigmatwo$ that can
be tuned) by an insulating layer that blocks tunneling of
electrons. The Cooper pairs in the JJA therefore interact with the
electrons in the 2DEG solely via the Coulomb interaction.}
\label{fig1}
\end{figure}

We begin with some simple observations on the effects of a
density-density coupling between the system of interest and a
diffusive 2DEG.  For concreteness, we consider the physical situation
shown in Figure
\ref{fig1}. 
A two-dimensional Josephson junction array (JJA) is separated by an
insulating slab from a two dimensional electron gas (2DEG), with
conductivity $\sigmatwo$. The Cooper pairs in the JJA couple to the
electrons in the 2DEG solely via the Coulomb interaction.  We assume
that the electron motion in the 2DEG is diffusive, and ignore weak
localization effects. For the experimental situation of interest this
is well justified since the typical 2DEG conductivities are large
enough that weak localization effects are insignificant at currently
accessible temperatures.

\subsection{Superconducting Phase}
Consider first the effect of the 2DEG on the properties of the
superconducting phase. In the absence of the 2DEG, the superconducting
state sustains a ``plasmon'' excitation associated with oscillations
of the phase of the superconducting order parameter. This mode is
gapless in two dimensions with energy $\sim \sqrt{k}$ at low momenta
$k$. This is in contrast to the linear dispersion of the sound mode
obtained in {\em neutral} superfluid systems ({\em i.e} with
short-ranged interactions). Upon coupling the charged superconductor
to the 2DEG, should we expect to recover the linear dispersing mode
due to the screening of the long-ranged part of the Coulomb
interaction? As shown below, the answer is no. On the time scale
relevant to the plasma oscillations, the slow diffusive motion of the
electrons in the 2DEG is unable to screen out the Coulomb interaction.
Thus the plasmon survives essentially unmodified.

To establish this result, note that in the superconducting state, the
phase of the Cooper pairs can be treated as a classical variable.  We
can then write down the classical equations of motion for the phase
$\phi(r,t)$ which obeys the Josephson relations:
\begin{eqnarray}
\label{eqnmn1}
\partial_t \phi(r,t) &=& 2 e U(r,t)\\
-2e\rho_s\nabla \phi(r,t) &=& \vec{J}_c
\end{eqnarray}
where $\vec{J}_s$ is the supercurrent, $U(r,t)$ is the potential and
$\rho_s$ is the superfluid stiffness (assumed constant). From charge
conservation in the JJA layer, we have ($\rho_c$ is the Cooper pair
density):
\be
\label{eqnmn2}
\partial_t \rho_c + \vec{\nabla} \cdot \vec{J}_c =0
\ee
The potential $U(r,t)$ is generated by both the Cooper pair charge
density as well as the electronic charge density in the 2DEG
($\rho_e$):
\begin{eqnarray}
\label{eqnmn3}
U(r,t) &=& \int d^2r'V(|r-r'|) (\rho_c(r',t)+\rho_e(r',t)) \\
\end{eqnarray}
where the interaction at large distances falls off as $V(|r-r'|)\sim
\frac1{|r-r'|}$ and, for convenience,  we have neglected the small but 
finite vertical separation between the JJA and the 2DEG.

Finally, to completely describe the dynamics of this system, we need
the equations of motions that govern charge motion in the 2DEG:
\begin{eqnarray}
\label{eqnmn4}
J_e(r,t) &=& - \sigmatwo \nabla U(r,t) - {\mathcal D} \nabla\rho_e\\ 0
&=& \partial_t \rho_e + \nabla\cdot J_e
\end{eqnarray}
where ${\mathcal D}$ is the diffusion constant for electrons, and
$J_e$ is the current density in the 2DEG. From equations
(\ref{eqnmn1},\ref{eqnmn2},\ref{eqnmn3},\ref{eqnmn4}), we obtain the
equation of motion for the phase variable, which is most conveniently
written in terms of its Fourier components $\tilde{\phi}(q,\omega)$
defined by:
\be
\tilde{\phi}(k,\omega) = \int d^2r dt e^{-ik\cdot r} e^{i\omega t} 
\phi(r,t).
\ee
The equation of motion is:
\be
[\frac{\omega^2}{(2e)^2}\frac1{\tilde{V}_{eff}} - \rho_s
k^2]\tilde{\phi}(k,\omega) = 0
\ee
where the effect of the 2DEG is simply to replace the Coulomb
interaction by the effective interaction $\tilde{V}_{eff}$:
\be
\frac1{\tilde{V}_{eff}(k,\omega)} = \frac1{\tilde{V}(k)} +\frac{\sigmatwo 
k^2}{-i\omega +{\mathcal D}k^2}
\ee
The Fourier transform of the $1/r$ Coulomb interaction in two
dimensions $\tilde{V}(k)$ takes the form
$\tilde{V}(k)=\frac{2\pi}{|k|}$ in the long wavelength limit.

In the absence of the 2DEG ($\sigmatwo=0$), we obtain the usual
two-dimensional plasmon mode:
\be
\omega_{pl}(k) = \sqrt{(2e)^2\rho_s\tilde{V}(k) k^2} =\sqrt{8\pi e^2\rho_s} 
\sqrt{k}
\ee 

Notice that with the 2DEG present, in the {\em static} limit
$\omega=0$, the Coulomb interaction is screened: \be
\tilde{V}_{eff}(k,\omega=0) = 1/(\frac{|k|}{2\pi} + \frac{\sigmatwo}{\mathcal 
D})
\label{screening}
\ee 
For situations that are not static, we cannot automatically assume
this screened Coulomb form for the interaction, rather, the dynamics
of the 2DEG will have to be taken into consideration.

The plasma mode is obtained by solving:
$$
\frac{\omega(k)^2}{(2e)^2}\frac1{\tilde{V}_{eff}(k,\omega(k))} - \rho_s 
k^2=0
$$
which, in the long wavelength ($k$ small) limit reduces to:
\be
\omega(k)= \omega_{pl}(k) - i(\pi\sigmatwo)k
\ee
Now, since the oscillation frequency of this mode goes as the square
root of the wave vector, it is much larger (in the long wavelength
limit) than the damping rate which depends linearly on wave vector
. Thus, as claimed above, the plasmon is a well defined excitation and
retains its $\sqrt{k}$ dispersion despite the presence of the 2DEG.

This problem of the phase dynamics deep inside the superconducting
state in the presence of a metallic gate, was examined earlier in
\cite{Gaitonde} where, in contast to the present work, the plasmon was 
found to be overdamped at long wavelengths. We believe this result to
be in error; in particular, the form of the dielectric function used
there (equation (20a) of reference \cite{Gaitonde}) is valid only in
three dimensions and not applicable to the two dimensional geometry of
the current problem.

\subsection{Nature of the Phase Transition}
\label{npt}
We now consider the nature of the zero temperature phase transition
between the superconducting and insulating states of the JJA. In this
section we present some general arguments that strongly suggest the
presence of a fixed line of critical points controlling the generic
(including disorder) superconductor-insulator transition of charged
particles in the presence of the metallic plane. Note that in the
presence of quenched disorder the superconductor to insulator
transition will be a continuous transition in two dimensions
\cite{Ma}.
 
Consider a system of bosons in two dimensions interacting via a $1/r$
Coulomb potential and coupled to a diffusive 2DEG in the manner
described. Such a system can be described in terms of a charged
bosonic field minimally coupled to the time component of a gauge field
($A_0$). The spatial components of this gauge field can be ignored
since we are in the extreme type II limit, at zero external field.
Here, we will examine the action governing the gauge field, which
takes the following form in imaginary time (a detailed derivation is
in the next section):
\begin{eqnarray}
S_A &=& \frac12\intk V_{eff}^{-1}(k,\omega)|\tilde{A}_0(k,\omega)|^2\\
\nonumber    &=&  \frac1{2e^{*2}} \intk 
\frac{|k|}{2\pi}|\tilde{A}_0(k,\omega)|^2 \\
             & &+ \frac{\sigmatwo}{2e^{*2}}\intk \frac{k^2}{|\omega|+D
k^2} |\tilde{A}_0(k,\omega)|^2
\label{sa}
\end{eqnarray}
where $e^*=2e$ is the charge of the Cooper pairs, and we have
introduced the compact notation $\dcross\omega =\frac{d\omega}{2\pi}$.

The first term on the right hand side of equation (\ref{sa}) gives
rise to the Coulomb interaction while the second term arises from the
coupling to the diffusive 2DEG. Notice that these terms contain
non-analytic functions of $k$,$\omega$. If we perfom a Renormalization
Group transformation on this system, by integrating out the boson and
gauge field modes that lie within an energy-momentum shell of finite
thickness, then additional terms of this form {\it cannot} be
generated. Hence the scaling of these operators arises solely from
rescaling of space-time and of the fields. Since the rescaling of the
gauge field is fixed by the requirement of gauge invariance, the
scaling dimensions of these operators can be easily derived. Gauge
invariance requires that the gauge field $A_0$ must scale as the
inverse time, to preserve the minimal coupling form of the
action. Hence its Fourier transform in two dimensions, $\tilde{A}_0$,
scales as:
\be
[\tilde{A}_0(k,\omega)] = {\bf L}^2
\ee   
where ${\bf L}$ has dimensions of length. Thus the Coulomb term has
dimensions:
\be
[\int d^2k d\omega |k||\tilde{A}_0(k,\omega)|^2]= {\bf L}^{1-z}
\ee
where $z$ is the dynamical critical exponent ($[$time$]={\bf
L}^z$). Thus, the coupling strength of the Coulomb interaction
$e^{*2}$ has the renormalization group flow:
\be
\label{eflow}
\frac{de^{*2}}{dln(b)}=e^{*2}(z-1)
\ee
when modes between $[\Lambda,\Lambda/b]$ are integrated out. Thus, a
fixed point that has a finite value of the Coulomb coupling
necessarily also has $z=1$\cite{Fisher}. This can also be seen from
the following simple argument. The $1/r$ Coulomb interaction in two
dimensions, is non-analytic in Fourier space. Therefore, the inverse
linear power law of the interaction cannot be renormalized. In
addition, if the fixed point has a finite value of the Coulomb charge,
then the energy must scale as $1/r$ which immediately implies that the
dynamical critical exponent is unity.

We can now apply a similar analysis to the second term in equation
(\ref{sa}). In the long wavelength limit, the ${\mathcal D}k^2$ term
in the denominator is irrelevant, so long as the dynamical critical
exponent for the theory satisfies $z<2$. This condition will need to
be checked at the end of the calculation to ensure consistency. Using
the known scaling dimension of the gauge field, we find the
interesting result:
\be
[\int d^2k d\omega \frac{k^2}{|\omega|}|\tilde{A}_0(k,\omega)|^2]=
{\bf L}^0
\label{scalingmarginal}
\ee
Not only is this operator marginal from its scaling dimension, it is
always strictly marginal! This is an exact statement, arising from
gauge invariance and the non-analytic form of the gauge field action,
and is {\it not} just the engineering (tree level) dimension of this
operator.

The presence of a strictly marginal operator in the fixed point action
will in general lead to a line of fixed points. This conclusion should
obtain if the renormalized Coulomb charge remains non-zero at the
transition.  Universal critical properties (such as exponents) will
vary continuously along this line, and hence will depend on
$\sigmatwo$.  As argued below, the Coulomb charge is expected to be
non-zero for the generic superconductor-insulator transition that
occurs in the presence of disorder, from a superconducting phase to a
gapless Bose-glass phase. This points to the existence of a fixed line
in the presence of the 2DEG.

In the absence of the metallic gate, the argument for a finite value
of the Coulomb charge at the superconductor to Bose-glass transition
was reported in \cite{Fisher}. There, the singularity in the long
wavelength behaviour of the compressibility (arising from the Coulomb
interaction) on both sides of the transition was used to argue that
the Coulomb charge remains finite at the transition itself. Although
this argument fails in the presence of the metal due to screening in
the static limit, a different argument can be made in this case to
obtain information about the generic transition. Imagine that at this
transition, the Coulomb charge does not take on a finite value, but
instead flows to zero. In this limit the gauge field will decouple
from the bosons, which then only experience short-ranged
interactions. The transition will therefore be in the dirty boson
universality class for which it is known that $z=2$
\cite{dirtybosons}. However, this scenario would not be consistent with 
the flow equation (\ref{eflow}) which indicates that near this fixed
point the Coulomb charge grows and renders it unstable. Hence we are
led to conclude that the Coulomb charge will remain non-zero at the
transition between the Bose-glass and the superconductor. This leads
to two possible scenarios. In the first scenario, the Coulomb charge
takes on a finite value at the transition, as in the case without the
metal. Then, a line of fixed points all with $z=1$ will obtain. In the
second scenario, the Coulomb charge flows away to infinity, and the
gauge fluctuations are completely controlled by the second term on the
right hand side of equation (\ref{sa}). Such fixed points could be
stable for $z>1$, and once again we are lead to expect a line of fixed
points, but with $z>1$. For consistency, we will also require $z<2$ to
justify our assumption in deriving (\ref{scalingmarginal}) where the
term containing the diffusivity which is quadratic in the momentum was
neglected in comparison to the frequency. The question of $z \ge 2$
fixed points in the presence of the metallic 2DEG will not be
discussed in the present work, and is left for future study.

Further support for a line of fixed points is provided by our
calculations on a model of the superconductor-insulator transition
with particle-hole symmetric disorder, where these scenarios are
explicitly realised. The critical properties of this model are
explored within a double epsilon expansion, and both a line of fixed
points with $z=1$, as well as a separate line of fixed points with
$z>1$ are found, corresponding to the two scenarios described above.

The situation in the clean problem, where the transition is between
the superconductor and the Mott insulator is, however, different. For
this special case, despite the presence of the strictly marginal
operator, none of the universal quantities are affected, since the
Coulomb charge flows to zero at the transition, which is hence
controlled by a single (3D XY) critical point. We will nevertheless
consider this problem first, since it will allow us to erect our
formalism in a technically more simple case. The properties of this
model are then considered within the large-$N$ and $\epsilon$
expansions. Subsequently, we consider a modified model, that includes
particle-hole symmetric disorder, which behaves more like the generic
case.  This model is studied using the $\epsilon$,$\epsilon_\tau$
expansion \cite{Cardy}\cite{Herbut} which reveals that indeed the
universal properties of the transition depend on the value of the 2DEG
conductivity $\sigmatwo$, and hence a fixed line results.

\section{Superconductor-Insulator Transition in the Clean Limit}
\label{sitcln}
In this section we describe a model of an array of Josephson junctions
coupled to a diffusive 2DEG via Coulomb interactions. We ignore the
presence of disorder and assume that the junctions are at integer
Cooper-pair filling.  (Note that in the clean limit, at non-integer
filling, the insulating phase would generally break translational
symmetry - a direct transition from the superconductor to such an
insulator is expected to be first order).

Consider an array of superconducting islands at positions labelled by
$\vec{r}$.  We model each of these islands as an $O(2)$ quantum rotor;
this is a legitimate approximation when there are on average many
Cooper pairs present on each island. The deviation of the boson charge
from the background value is denoted by $n_r$. Then, the number of
Cooper pairs on each site $\frac{n_r}{e^*}$ is conjugate to the phase
on each island $\phi_r$.
\be
[\frac{n_r}{e^*},e^{i\phi_r}] = e^{i\phi_r}
\ee
Coupling the charge fluctuations in the JJA layer to the diffusive
2DEG via Coulomb interactions as described earlier, we obtain:
\begin{eqnarray}
\nonumber
\hat{H} &=&-E_J\sum_{<rr'>} \cos(\phi_r -\phi_{r'}) \\
\nonumber
& &+ \frac12 \sum_{rr'}(n_r+\delta n_e(r))V(r-r')(n_{r'}+\delta
n_e(r'))\\ & & + H^0_{2DEG}
\end{eqnarray}
where $\delta n_e$ is the fluctuation of electronic charge density in
the 2DEG, the Coulomb interaction at large distances is taken to be
$V(r-r')\sim 1/2\pi|r-r'|$ (the factor of $2\pi$ is introduced for
later convenience). $ H^0_{2DEG}$ is the Hamiltonian of the diffusive
2DEG, without the Coulomb interaction between electrons. We now recast
this as a path intergal over the bosonic ($n$,$\phi$) and fermionic
($\psi_e$) degrees of freedom. It is also convenient to introduce a
Hubbard-Stratonovich decoupling of the long-range interaction using an
auxilliary field $A_0$.  (As will be clear, $A_0$ may be interpreted
as the scalar component of a gauge field.)  This leads to the
following representation for the finite temperature ($=1/\beta$)
partition function:
\be
{\mathcal Z}= \int {\mathcal D}\phi {\mathcal D} n{\mathcal
D}A_0{\mathcal D}\psi_e e^{-S}
\ee
where the phase satisfies the boundary conditions
$\phi_r(\beta)=\phi_r(0)+2\pi m_r$ with $m_r$ any integer, and
\begin{eqnarray}
S &=&S_B + S_{int} + S_A[A_0] + S^0_{2DEG}[\psi_e] \\
\nonumber
S_B &=& \int_\tau [\frac{i}{e^*}\sum_r
n_r(\tau)\partial_\tau\phi_r(\tau)
\\
    & & -E_J\sum_{<rr'>}\cos(\phi_r(\tau)-\phi_{r'}(\tau))]\\ S_{int}
&=& \frac{i}{e^*}\int_\tau \sum_r A_0(r,\tau)(n_r(\tau)+\delta
n_e(r,\tau))\\ S_A &=& \frac1{2e^{*2}} \int_\tau \sum_{rr'}A_0(r,\tau)
V^{-1}(r-r') A_0(r',\tau).
\end{eqnarray}
Here $S^0_{2DEG}$ is the action correspoding to the diffusive 2DEG in
the absence of the Coulomb interaction, $\int_\tau=\int_0^\beta d\tau$
and $V^{-1}$ implies the matrix inverse of the potential.  Following
the pioneering work of Hertz\cite{Hertz} on quantum phase transitions
in fermionic systems, we integrate out the diffusive electrons in the
2DEG to derive an effective action for the bosons. The contribution
arising from this proceedure is:
\begin{eqnarray}
{\mathcal Z}[A_0] &=& \int {\mathcal D}\psi_e e^{-S^0_{2DEG}}e^{-\frac
i{e^*}\int_\tau\sum_r A_0(n_r(\tau)+\delta n(r,\tau))}\\ &=& {\mathcal
Z}[0] <e^{\frac{i}{e^*}\int_\tau\sum_r A_0 (n+\delta n)}>_{2DEG}
\end{eqnarray}
Performing a cumulant expansion for the average, and taking into
account $<\delta n(r,\tau)>_{2DEG}=0$ we have:
\bea
\nonumber
&&<\exp{\frac{i}{e^{*}}\int\sum_r A_0 (n+\delta n)}>_{S^0_{2DEG}} =
\\\nonumber
\exp &-&\frac1{2e^{*2}}\int_{\tau,\tau'}\sum_{r,r'}<\delta n_r(\tau)\delta 
n_{r'}(\tau')>_{S^0_{2DEG}}A_0(r,\tau)A_0(r',\tau')\\ &+& ...
\eea
 As in Ref. \cite{Hertz}, we shall neglect the three and higher body
interaction terms that are generated to obtain:
\be
{\mathcal Z}[A_0] \cong
e^{-\frac12\int_{\tau,\tau'}\sum_{r,r'}\Pi^0(r-r',\tau-\tau')A_0(r,\tau)A_0(r',\tau')}
\ee
where
\be
\Pi^0(r-r',\tau-\tau') = \frac1{e^{*2}}<\delta n_r(\tau)\delta 
n_{r'}(\tau')>_{2DEG}.
\ee
We thus obtain:
\begin{eqnarray}
\nonumber
{\mathcal Z} & = & \int {\mathcal D}\phi {\mathcal D} n{\mathcal
D}A_0{\mathcal D} e^{-S_{eff}} \\ S_{eff} &=& \int _\tau
\frac{i}{e^*}\sum_r n_r(\partial_\tau \phi + A_0) -
E_J\sum_{rr'}\cos(\phi_r - \phi_r')\\ & &
+\frac1{2e^{*2}}\int_{\tau\tau'}A_0(r,\tau)
V_{eff}^{-1}(r-r',\tau-\tau')A_0(r',\tau')
\end{eqnarray}
where
\be
V_{eff}^{-1}(r,\tau) = V^{-1}(r)\delta(\tau) + \Pi^0(r,\tau).
\ee
For a diffusive metal the Fourier transform of $\Pi^0$ is given by:
\be
\tilde{\Pi}^0(k,\omega) = \sigmatwo \frac{k^2}{|\omega|+D k^2}
\ee
where $\sigmatwo$ is the conductivity and ${\mathcal D}$ is the
diffusivity of the metal.  Thus, the Fourier transform of the
effective potential is:
\bea
\frac1{\tilde{V}_{eff}(k,\omega)} &=& \frac1{\tilde{V}(k)} + \sigmatwo 
\frac{k^2}{|\omega|+D k^2}\\
				  &=& |k| + \sigmatwo
\frac{k^2}{|\omega|+D k^2}
\eea
Note that in the static ($\omega=0$) limit, the Coulomb interaction is
screened as can be seen in equation (\ref{screening}). However, the
question of screening is more subtle when it comes to dynamical
phenomena (as we have seen in the case of the plasmon).

\subsection{Shift in Phase Boundary}
In this section we study the dependence of the zero temperature
superconductor-insulator phase transition point on the 2DEG
conductivity.  Let us call the critical Josephson coupling $E_J^c$;
for $E_J>E_J^c$ the system is in the superconducting phase. We would
like to find the dependence of $E^c_J$ on the 2DEG parameters. A rough
estimate of this phase boundary is obtained by starting out in the
superconducting phase and applying an analogue of the Lindemann
criterion; i.e. we ask for what value of the parameters are the
superconductor phase fluctuations of O(1).

It is convenient for this analysis to rewrite the partition function
for the sytem in terms of the phase variable:
\bea
\nonumber
{\mathcal Z} = \int {\mathcal D}\phi \exp &-&\frac12\intk
\frac{\omega^2}{e^{*2}}[\tilde{V}^{-1}_{eff}(k,\omega)]|\tilde{\phi}_{k\omega}|^2 
\\
	&+&E_J\int_\tau \sum_{rr'}\cos(\phi_r-\phi_{r'})
\label{effaction}
\eea
with the bounary conditions $\phi_r(\beta)=\phi_r(0)+2\pi m_r$, where
$m_r$ is integer. In the superconducting state, we can replace the
cosine in the action (\ref{effaction}) by a quadratic gradient
term. The quadratic action for the phase fluctuations in the
superconductor is:
\be
S_{sc} = \intk [\frac{\omega^2}{e^{*2}} \tilde{V}^{-1}_{eff}(k,\omega)
+ (E_Ja^2)k^2]|\tilde{\phi}_{k\omega}|
\label{Ssc}
\ee
where $a$ is the microscopic cutoff, and the integral over modes in
momentum space is restricted to $|k|<\sqrt{4\pi}/a$. This combined
with the `Lindemann' condition (${\mathcal A}$ is a constant of O(1)):
$$
<\phi^2>={\mathcal A} \cong O(1)
$$
leads to the follwing implicit equation for the critical point
$E^c_J$:
\be
{\mathcal A}=\intk
(\frac{\omega^2}{e^{*2}}[|k|+\sigmatwo\frac{k^2}{|\omega|+{\mathcal
D}k^2}]+E_Ja^2k^2)^{-1}
\ee

\begin{figure}
\epsfxsize=2.8in
\centerline{\epsffile{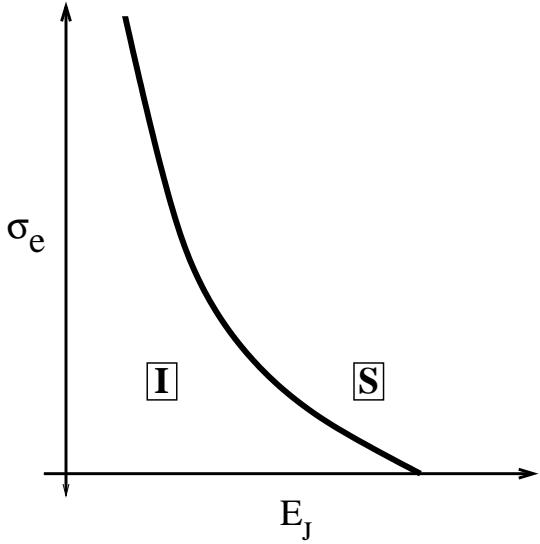}}
\vspace{0.1in}
\caption{The Superconductor-Insulator phase boundary in the presence of a 
2DEG (conductivity $\sigmatwo$). The 2DEG stabilizes the
superconductor, as described in the text. Thus, the critical value of
the intergrain Josephson coupling $E_J$ at which superconductivity is
established, decreases with increasing $\sigmatwo$.}
\label{phasebdy}
\end{figure}

Inspecting the integral it is clear that the superconductor is
stabilized in the presence of the 2DEG, i.e. the phase transition
point $E^c_J(\sigmatwo)$ is reduced compared to its value in the
absence of the 2DEG: $E^c_J(\sigmatwo)<E^c_J(0)$. This is
schematically depicted in Figure (\ref{phasebdy}). Qualitatively, this
effect is to be expected, since the screening arising from the 2DEG
contributes to weakening the Coulomb interaction, and hence enhances
the stability of the superconduting state. Also, it has been argued
\cite{Kivelson} that coupling the phase fluctuations to external
degrees of freedom (`dissipation') leads to reduced phase fluctuations
which would also stabilize the superconducting phase. This effect has
been observed in the experiments of Rimberg et al.\cite{Rimberg} who
were able to cross the insulator to superconductor phase boundary just
by increasing the conductivity of the 2DEG. Experiments on Josephson
junction arrays with chromium shunt resistors (that do not go
superconducting), have been reported in \cite{japanesegroup}. A
phenomenological model of that system can be written down which, in
the superconducting phase, would be exactly the same action as in
equation (\ref{Ssc}) above without the ${\mathcal D}k^2$ term, with
the conductance $\sigmatwo$ being the conductance of the chromium
resistor.  In that experiment too, it was found that increasing the
value of $\sigmatwo$ increases the stability of the superconducting
region.

\subsection{Critical Properties: Clean Case}
In order to study the critical properties of this system, it is
convenient to reformulate the action in equation (\ref{effaction}), by
rewriting it as a ``soft spin'' model, where, after coarse graining,
the rotor ``spin'' $e^{-i\phi}$ is replaced by a complex field
$\psi$. This automatically takes into account the non-trivial boundary
conditions on the phase $\phi(\beta)$. The finite temperature
partition function can then be written as a path integral over these
fields:

\begin{eqnarray}
\nonumber 
{\mathcal Z} &=& \int {\mathcal D}\psi {\mathcal D}A_0 e^{-S_\psi -
S^{eff}_A}\\
\nonumber
S_\psi &=& \int d^2r d\tau \frac1c |(\partial_\tau-iA_0)\psi|^2 +
c|\nabla\psi|^2 + r|\psi|^2 + \frac{u}4|\psi|^4\\ S^{eff}_A &=&
\frac1{2e^{*2}}\intk[|k|+\sigma_e\frac{k^2}{|\omega|+{\mathcal 
D}k^2}]|A_0(k,\omega)|^2
\label{Saeff}
\end{eqnarray}
Since we are interested in the critical properties of this model, we
can safely ignore the ${\mathcal D}k^2$ part in the denominator of the
second term in (\ref{Saeff}). Also, we will scale out the bare sound
wave velocity $c$ that appears in the action above by a suitable
rescaling of the imaginary time. All velocities will therefore be
measured in units of this bare velocity. For convenience, we continue
to use the same symbols for the scaled quantities, but these factors
will have to be restored when comparing with the experiments.

In this section we consider the effect of the 2DEG and the Coulomb
interaction on the properties of the superconductor-insulator
transition in the {\it absence} of disorder. This is studied within
two different approximations - the large-$N$ technique and the
$\epsilon$ expansion.  As shown below, in both these approximation
schemes, the Coulomb interaction is found to be marginally irrelevant,
even in the presence of the metal. As a result, the gauge field
decouples from the bosonic degrees of freedom at the critical point,
and the presence of the strictly marginal operator (the coupling to
the 2DEG) does not affect the critical properties of the system. The
transition is therefore controlled by the same fixed point as in the
clean case with short-range interaction and without any coupling to
the metal. The agreement between the large-$N$ and $\epsilon$
expansions is strong evidence that neither the Coulomb interaction or
the coupling to the $2DEG$ affects the critical properties in the
clean commensurate superconductor-insulator transition.

In the generic case, away from this idealized limit, we however expect
that the combination of the Coulomb interaction and the coupling to
the $2DEG$ will lead to a fixed line controlling the transition. This
is illustrated in the next section by introducing disorder into the
model.

\subsubsection{Large N Calculation:}
\label{section2D}
The effect of the 2DEG on the critical properties of the clean problem
can be considered within the large-$N$ approximation. We remain in
$D=2$ spatial dimensions, but generalize the model of a single species
of boson to one that has $N$ flavours of complex boson fields
$\psi_i$:
\begin{eqnarray}
\nonumber
S_{\psi}&=& \int d^2rd\tau \sum_{i=1}^{N} |(\partial_\tau -
iA_0)\psi_i|^2 + |\nabla\psi_i|^2 + r_0|\psi_i|^2 + \\ & &\frac{u}{N}
[\sum_i |\psi_i|^2]^2 \\ S_A &=& \frac{N}{2e^{*2}} \intk
[|k|+\sigmatwo\frac{k^2}{|\omega|}]|\tilde{A}_0(k,\omega)|^2
\label{SlargeN}
\end{eqnarray}

As argued previously, the entire renormalization of the Coulomb charge
$e^*$ comes from rescaling the fields and the momenta. This leads to
the following exact equation:
\be
\frac{de^{*2}}{d\ln b} = (z-1)e^{*2}
\label{flow}
\ee 
while the term containing $\sigmatwo$ is exactly marginal. Here, we
will derive an expression for the dynamical exponent ($z$) at the
critical point - within a large-$N$ approximation.

In the $N\rightarrow \infty$ limit, the (saddle point) solution to the
above problem has $A_0=0$, and the critical point, at this order, is
identical to that of the short-ranged model. We therefore need to go
to next order in $1/N$ to see the effect of the gauge field coupling
. In calculating $z$ to order $1/N$ at the critical point, we only
need to consider the self energy diagram shown in figure
(\ref{largeN}a) but within this approximation the gauge propagator
${\mathcal G}_A$ is replaced by its RPA form:
\be
\frac1N {\mathcal G}^{-1}_A(k,\omega) = e^{*-2}[|k| + \sigmatwo 
\frac{k^2}{|\omega|}]+\frac{k^2}{16\sqrt{k^2+\omega^2}}
\ee
as shown in Figure \ref{largeN}c. Therefore we find the dynamical
exponent is given by:
\begin{eqnarray}
z-1 &=& \frac1N I(e^{*2},\sigmatwo)\\\nonumber I(e^{*2},\sigmatwo)&=&
\int_0^{\pi/2} \frac{d\theta}{2\pi^2}
\frac{\sin^2\theta(1-3\sin^2\theta)}{e^{*-2}(1+\sigmatwo\tan\theta)+\frac{\sin\theta}{16}}
\end{eqnarray}
(Details of this derivation are relegated to Appendix \ref{appendixa}). In the absence of the metal and for small Coulomb
charge we find: $I(e^*\rightarrow
0,\sigmatwo=0)=-\frac5{32\pi}e^{*2}$, and implies that a weak Coulomb
interaction is marginally irrelevant at the short-ranged fixed point
in the absence of the metal.  Even in the presence of the metal, it is
found that $I(e^{*2},\sigmatwo)<0$, and hence the Coulomb interaction
continues to be marginally irrelevent in the large-$N$
approximation. Thus, the only fixed point solution to the flow
equation (\ref{flow}) is $e_c^{*}=0$. At this fixed point, the gauge
field coupling may be ignored and the presence of the metal does not
affect any of the universal critical properties.  Therefore in the
large-$N$ approximation despite the presence of the metal, the
transition is in the universality class of the short-ranged model.

\subsubsection{The $\epsilon$ Expansion for the Clean Problem}  
In the dimensionality of interest $D=2$, the quartic term $|\psi|^4$
is relevant by power counting. One way to control the RG flows is via
the standard technique of considering the problem close to its upper
critical dimension, the dimension at which the quartic coupling is
also marginal by power counting ($D=3$). Therefore, we write the
effective action (\ref{Saeff}) in $D=3-\epsilon$ dimensions as:
\begin{eqnarray}
\label{seffpsi}
S &=& S_\psi + S_A\\ S_\psi &=& \int d^Drd\tau
|(\partial_t-iA_0)\psi|^2 + |\nabla\psi|^2 + r|\psi|^2 +
\frac{u}4|\psi|^4\\ S_A &=& \frac1{2e^{*2}}\int \dcross^Dk
\dcross\omega (|k|^{D-1} +\sigmatwo
\frac{k^D}{|\omega|})|\tilde{A}_0(k,\omega)|^2
\label{noD}
\end{eqnarray}
Notice that the action for the gauge field is written in a general
dimension $D$, so as to always produce the $1/r$ form of the Coulomb
interaction (if $\sigmatwo=0$). The term containing $\sigmatwo$ is
continued to general dimensions so that it remains an exactly marginal
operator.

This problem, in the absence of the metal ($\sigmatwo=0$) was
considered in \cite{fishercoulomb} where the Coulomb interaction was
found to be marginally irrelevant. Here too we will investigate the
flow of the Coulomb coupling, but in the presence of the metal. For
$D<3$, the non-analyticity of the Coulomb term in momentum space once
again ensures that it is not renormalized by terms that are generated
on integrating out high energy modes. Thus equation (\ref{flow}) again
obtains for the flow of $e^*$ while the term containing $\sigmatwo$ is
exactly marginal. Here, we will calculate $z$ to first order in
$\epsilon$, the relevant diagram is shown in Figure
\ref{figselfenergy}a. We find (see Appendix \ref{sectionselfenergy} for details):
\begin{eqnarray}
z-1 &=& -\frac{e^{*2}}{6\pi^2}A(\sigmatwo)\\ A(\sigmatwo)&=& \frac2\pi
\int^{\pi/2}_0 d\theta
\frac{\sin^2\theta(4\sin^2\theta -1)}{1+\sigmatwo\tan\theta}
\end{eqnarray}
where $A(\sigmatwo)$ is normalized so that $A(0)=1$.  In that limit,
$\sigmatwo=0$, we recover the flow equation of
Ref. \cite{fishercoulomb}, so that the Coulomb interaction is
marginally irrelevant since the right hand side of (\ref{flow}) is
negative.  Even for non zero $\sigmatwo$, it can be shown that
$A(\sigmatwo)>0$ and thus the Coulomb interaction is marginally
irrelevant despite the presence of the metal.  The fixed point value
of the Coulomb interaction then is $e_c^{*}=0$. As a result, the gauge
field is decoupled from the bosons at the critical point, and the
presence of the strictly marginal operator does not affect the
universal critical properties.

\section{Line of Fixed Points in a Model with Disorder}
In this section, we present evidence to support our claim that the
critical properties for the generic transition (as opposed to the idealized clean commensurate
limit in Section \ref{sitcln} above)
are controlled by  a line of fixed points. We study
a model of the disordered superconductor-insulator transition in the
presence of Coulomb interactions and a metallic plane.  This is in
contrast to the zero disorder transition considered above. (Note that
even in the absence of the metal the clean commensurate
superconductor-insulator transition with Coulomb interactions has
behaviour different from the generic case). We consider instead a
slightly more realistic model that contains a special form of disorder
that preserves the particle hole symmetry of the problem. In the
absence of the metal, such a model was recently studied by
Herbut\cite{Herbut} who found a $z = 1$ fixed point with a finite
value of the Coulomb charge. Here we add the coupling to the metal,
and first demonstrate that {\em a line of fixed points} with a finite
value of the Coulomb charge and $z=1$ can be obtained. Subsequently, a
line of fixed points with $2>z>1$, for which the Coulomb charge flows
to infinity, is also obtained.

The model we consider is essentially the one in equation
(\ref{seffpsi}) but with a coupling to disorder $\mathcal V$,
included:
\begin{eqnarray}
\label{dis}
S &=& S_\psi + S_A + S_{dis}\\\nonumber S_\psi &=& \int d^Drd\tau
(|(\partial_t-iA_0)\psi|^2 + |\nabla\psi|^2 + r|\psi|^2 + \frac
u4|\psi|^4)\\\nonumber S_A &=& \frac1{2e^{*2}}\int \dcross^Dk
\dcross\omega (|k|^{D-1} +\sigmatwo
\frac{k^D}{|\omega|})|\tilde{A}_0(k,\omega)|^2\\\nonumber S_{dis} &=&
\int d^Dr d\tau {\mathcal V}(r) |\psi(r,\tau)|^2
\end{eqnarray}
Notice, that the coupling to disorder preserves the particle hole
symmetry of the model (the action is invariant under the particle-hole
transformation: $\psi\rightarrow\psi^*$, $A_0\rightarrow -A_0$). For
quenched disorder, the random potential ${\mathcal V}$ is independent
of imaginary time, and therefore has a big effect on the physics of
the system. In the following, it is convenient to generalize it to a
random variable that is correlated in $\epsilon_\tau$ dimensions; in
other words we assume the following statistics for the random
potential:
\begin{eqnarray}
<{\mathcal V}> &=& 0\\ <{\mathcal V(R)}{\mathcal V(R')}> &=& \bar{W}
\delta^{D+1-\epsilon_\tau}(R-R')
\end{eqnarray}
where the label $R$ runs over both space and time components. For
quenched disorder $\epsilon_\tau=1$. However, it will be convenient to
study the model in the limit of small $\epsilon_\tau$ which will allow
one to control the effects of disorder. This approximation was
introduced in Ref. \cite{Cardy} to study the problem of bosons in the
presence of disorder and short-ranged interactions.

In order to perform the disorder average, we use the standard replica
technique. We introduce $n$ copies of the fields
$\psi\rightarrow\psi_\alpha$, $\alpha=1\dots n$, average over
different realizations of the disorder potential, and finally take the
$n\rightarrow 0$ limit. Averaging over disorder generates the
following term which has a non-local interaction in $\epsilon_\tau$
dimensions between the replica fields.
\bea
\label{sdis}
S &=& \sum_{\alpha=1}^n S_B[\psi_\alpha] + S_A + S'_{dis}\\
\nonumber
S'_{dis} &=& -\frac{\bar{W}}2 \sum_{\alpha,\beta=1}^n \int d^{D+1}R
d^{D+1}R'\delta^{D+1-\epsilon_\tau}(R-R')\\ &
&|\psi_\alpha(R)|^2|\psi_\beta(R')|^2
\eea
We now consider performing an RG transformation on the above action,
by integrating out the short wavelength modes $1/\Lambda
<a<b/\Lambda$, and then rescaling momenta, frequency and fields to
obtain the flow equations for the various couplings.  The calculation
is controlled by working in $D=3-\epsilon$ space dimensions, and as
mentioned, with disorder that is correlated in $\epsilon_\tau$
dimensions.  We assume that $\epsilon$,$\epsilon_\tau$ are small and
derive the flow equations to lowest order in these quantities. Details
of this derivation are relegated to Appendix \ref{appendixb}. After a redefinition of the coupling constants:
\begin{eqnarray}
\nonumber
q^2 &=&\frac{e^{*2}}{2\pi^2}\\ \nonumber
\lambda &=& \frac{U}{8\pi^2}\\
W &=& \frac{\bar{W}}{2\pi^2}
\label{couplings}
\end{eqnarray}
the flow equations take the following form:
\bea
\label{coulomb}
\frac{dq^2}{d\ln b}&=& (z-1)q^2 = (\frac W8 -\frac{A}3 q^2)q^2\\
\label{lambda}
\frac{d\lambda}{d\ln b}&=& \epsilon \lambda + \frac{B}2\lambda q^2 
-\frac52 \lambda^2 -\frac{C}4q^4 + \frac{11}8\lambda W\\
\label{W}
\frac{dW}{d\ln b}&=& (\epsilon + \epsilon_\tau)W + \frac78 W^2 + \frac{B}2 
q^2 W -2 \lambda W
\eea
  and the coupling constant $\frac1g = \frac{\sigmatwo}{e^{*2}}$ is
strictly marginal and does not flow. The parameters $A$,$B$,$C$ that
appear in the flow equations are functions of $\sigmatwo$ and have
been normalized so that $A(0)=B(0)=C(0)=1$. They are given by the
following expression:
\bea
A(\sigmatwo) &=& \frac2\pi \int^{\pi/2}_0 d\theta
\sin^2\theta(4\sin^2\theta -1)f(\sigmatwo,\theta)\\
B(\sigmatwo) &=& \frac4\pi \int^{\pi/2}_0 d\theta \sin^2\theta
f(\sigmatwo,\theta)\\ C(\sigmatwo) &=& \frac4\pi \int^{\pi/2}_0
d\theta \sin^2\theta f^2(\sigmatwo,\theta)
\eea
where
\be
f(\sigmatwo,\theta) = \frac1{1+\sigmatwo\tan \theta}
\label{g}
\ee
.

\begin{figure}
\epsfxsize=2.8in
\centerline{\epsffile{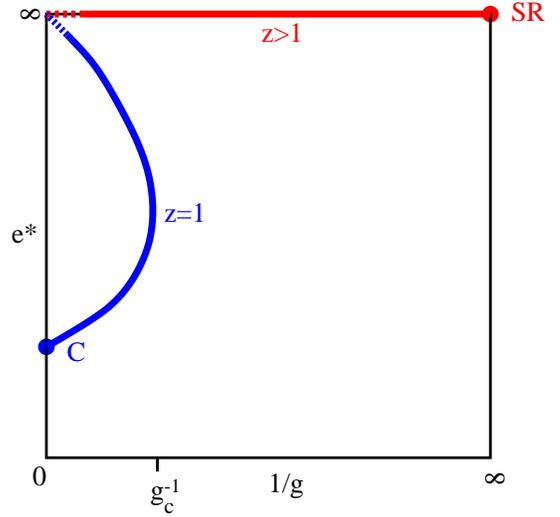}}
\vspace{0.1in}
\caption{
The results of the $\epsilon$, $\epsilon_\tau$ expansion showing the
fixed lines obtained in the presence of the marginal coupling $1/g$
which is proportional to the 2DEG conductivity. Only the fixed point
value of the Coulomb charge $e^*$ is plotted. The coupling constants
$e^*$ and $g$ are measured in units of $\epsilon$, where
$\epsilon_\tau=\epsilon$ has been assumed. The $z=1$ fixed points are
obtained for $g^{-1}<g_c^{-1}$, while $z>1$ fixed points are obtained
for all $g$. The fixed points with both $e^*$ and $g$ large are not to
be trusted since gauge field fluctuations are big in that case, and
are shown with dashed lines. The limiting cases shown as C and SR,
correspond to the Coulomb fixed point found in [13]
and the short-ranged fixed point found in [12]
}
\label{exp}
\end{figure}

\subsection{Fixed line with $z=1$}
We now solve for the fixed points of the flow equations above,
assuming that the Coulomb charge $e^*$ takes on a finite value at the
transition. The coupling contants at the fixed point are proportional
to $\epsilon$ but depend both on the ratio $\epsilon_\tau/\epsilon$
and on $\sigmatwo$. In the following we will assume
$\epsilon_\tau/\epsilon=1$, a choice that is consistent with our aim
to study the $\epsilon_\tau=\epsilon=1$ limit. We evaluate the
functions $A$,$B$,$C$ numerically for various values of $\sigmatwo$
and find the fixed point values of the couplings which are quantities
of order $\epsilon$ that depend on the value of $\sigmatwo$. It turns
out that fixed point solutions of this kind exist for $\sigmatwo$ in
the range $0<\sigmatwo<1.99$. While it is convenient to solve for the
fixed points in terms of $\sigmatwo$, they are more appropriately
labelled by the value of the marginal coupling $g$, which can be
obtained from the relation: $\frac1g =
\frac{\sigmatwo}{2\pi^2 q_c^2}$. When expressed in terms of this
variable, the fixed points exist in the range $\frac1g<\frac1{g_c}
\approx (572 \epsilon)^{-1}$. In fact, as shown in figure \ref{exp}, for
every value of $g$ in this range, there are a pair of fixed
points. The fixed points with a smaller value of the Coulomb charge,
evolve in the limit $\frac1g\rightarrow0$ (which corresponds to
removing the metallic plane) to the fixed point studied by Herbut in
\cite{Herbut}.

All of the fixed points obtained are found to be stable within a
linear stability analysis; although the eigevalues are found to be
complex, they have negative real parts. Thus we have a stable line of
fixed points for a range of the 2DEG conductivity. The dynamical
critical exponent $z$ for all of these fixed points is unity. Other
critical exponents can be expressed in terms of the fixed point
couplings:
\be
\eta = \epsilon q_c^2(\sigmatwo) B(\sigmatwo)
\ee
while the exponent $\nu$ can be expressed as:
\be
\nu = \frac12 
+\frac14(\lambda_c(\sigmatwo)-\frac{D(\sigmatwo)}{12}q^2_c(\sigmatwo)-\frac{W_c(\sigmatwo)}4)
\label{nu}
\ee
where the function $D(\sigmatwo)$ (which has been normalized to be
unity in the absence of the 2DEG, $D(0)=1$) is given by:
\be
D(\sigmatwo) = \frac 4\pi\int^{\pi/2}_0 d\theta \sin^2 2\theta
f(\sigmatwo,\theta)
\label{D}
\ee

\begin{figure}
\epsfxsize=2.8in
\centerline{\epsffile{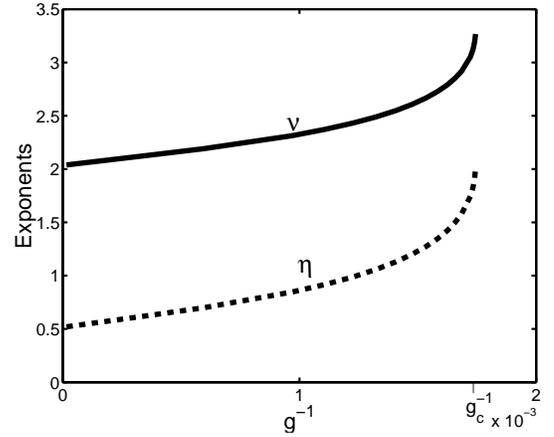}}
\vspace{0.1in}
\caption{The exponents $\eta$ and $\nu$ for the $z=1$ fixed line (lower 
branch in Figure 3) controlling the S-I phase transition. Along the
horizontal axis is plotted the marginal coupling $1/g$ which is
proportional to the 2DEG conductivity. For this plot we have set
$\epsilon=\epsilon_\tau=1$. From the dependence of the exponents on
the conductivity it is clear that the critical properties are
significantly affected by the presence of the 2DEG.}
\label{exponents}
\end{figure}

The remaining exponents can all be derived from a knowledge of these,
if hyperscaling is assumed. The Chayes inequality $\nu \geq 2/D$ is
satisfied by all these fixed points. A plot of these two critical
exponents as a function of the marginal coupling $1/g$ (which is
proportional to the 2DEG conductivity) is shown in Figure
(\ref{exponents}). Only the exponents for the lower branch of the
$z=1$ fixed line in figure \ref{exp} are shown.  The exponent $\nu$ is
found to increase with increasing 2DEG conductivity.  If this
qualitative effect that is present in the $\epsilon$ expansion
persists unchanged in the physical problem, then this would predict
that the critical region for a finite sample would increase as the
2DEG conductivity increases, which could be experimentally
accessible. We return to this point in the following section.
 
This calculation therefore, provides us with an explicit demonstration
of the general claims of this paper - that in the presence of the
Coulomb interaction and a metal, the dirty superconductor to insulator
transition can be controlled controlled by a line of fixed points,
parametrized by the metal conductivity.

\subsection{Fixed Line With $z>1$}
We now turn to the question of whether there are any stable fixed
points in this model with $e^*=\infty$. For clarity we rewrite the
gauge action as:
\begin{eqnarray}
\nonumber
S_A &=& \frac1{2e^{*2}} \int \dcross^Dk \dcross\omega \,
|k|^{D-1}|\tilde{A}_0(k,\omega)|^2 \\ & & +\frac1{2g} \int \dcross^Dk
\dcross\omega\,
\frac{k^D}{|\omega|}|\tilde{A}_0(k,\omega)|^2
\label{denominator}
\end{eqnarray}
where again $g$ has beens used to denote the coupling strength of the
marginal term, and is related to the bare couplings by
$g=e^{*2}/\sigmatwo$. Under the Renormalization Group transformation,
$g$ does not flow. Now, if the renormalized Coulomb charge were to
flow to infinite strength, then the gauge fluctuations would entirely
be controlled by the second term in the gauge action in equation
(\ref{denominator}). In order that this is a stable fixed point we
need to ensure that $z>1$, so that $\beta(e^*)>0$ at the fixed
point. In this limit, the flow equations can be obtained from the
expressions (\ref{coulomb} - \ref{g}) by formally taking the limit
$e^*$, $\sigmatwo\rightarrow\infty$ but holding the ratio
$e^{*2}/\sigmatwo$ finite and equal to $g$. Defining:
$$
\bar{g}=\frac g{2\pi^2} 
$$
yields the following flow equations:
\bea
\label{lambdaeinfinity}
\frac{d\lambda}{d\ln b} &=& \epsilon \lambda + \frac1\pi\lambda \bar{g} 
-\frac52 \lambda^2 -\frac14\bar{g}^2 + \frac{11}8\lambda W\\
\label{Weinfinity}
\frac{dW}{d\ln b}&=& (\epsilon + \epsilon_\tau)W + \frac78 W^2 + \frac1\pi 
\bar{g} W -2 \lambda W 
\eea
while
\be
z-1 = \frac{W}8 -\frac{\bar{g}}{3\pi}
\ee
Assuming that $\bar{g}$ is a constant of order $\epsilon$, and working
with $\epsilon=\epsilon_\tau$, we can solve for the fixed points in
this theory to first order in epsilon, as a function of the marginal
coupling $g$. Fixed points are found to exist for all such values of
$g$, and they are all found to have $z>1$. A linear stability analysis
of these fixed points demonstrates that they are stable (eigenvalues
of the stability matrix are found to be complex, but with negative
real parts). Thus, a stable fixed line of critical points with $z>1$
can be obtained within this model.


In order to derive this model we have assumed that the dynamic critcal
exponent satisfies $z<2$. This allows us to retain just the frequency
part in the denominator in passing from equation (\ref{Saeff}) to
equation (\ref{noD}), which is the form used in all the subsequent
analysis. In the epsilon expansion, the dynamical critical exponent
differs from unity by an amount of order $\epsilon$. Since $\epsilon$
is considered the smallest quantity in the problem, the condition on
the dynamical critical exponent $z<2$ is formally satisfied within our
calculation.

These results are summarized in Figure \ref{exp}, where only the
couplings $e^*$ and $g$ for the fixed points is plotted. The $z>1$
fixed points are found for all values of $1/g$ (which is proportional
to the conductivity of the 2DEG). However, for very small values of
$1/g$ they are not to be trusted, since we assume $g$ to be
$O(\epsilon)$. Therefore, these fixed points are shown with dashed
lines. In the limit of very large 2DEG conductivity ($1/g$ large), the
fixed point obtained is the same as for the problem with short-ranged
interactions and no metallic plane, which was considered in
\cite{Cardy} (this is made explicit in the Appendix \ref{appendixd}).

\section{Quantum Hall transitions with a metallic plane}

In this section we consider the effect of a metallic plane on the quantum Hall transition, which, along with the
superconductor to insulator transition, has been one of the most
intensively studied quantum critical phenomena\cite{generalrefs}. Most
theoretical studies of the quantum Hall transition between integer
quantum Hall states have used a non-interacting model of electrons in
a random potential subject to a magnetic field. While it is believed
that short-ranged interactions between the electrons are
irrelevant\cite{dhlee}, the same is not true of the long-ranged
Coulomb interaction which is present in the physical problem of
interest. It has been suggested that the non-interacting electron
universality class for these quantum Hall transitions may be partially
recovered by screening of the long-range Coulomb interactions by a
nearby metallic gate.  Increasing the screening in this way was shown
experimentally to induce a change of variable-range hopping transport,
away from the transition, from the Efros-Shklovksii behavior expected
with Coulomb interactions to the Mott behavior expected for
noninteracting electrons.~\cite{screening} So one conjecture would be
that a nearby metallic gate simply changes the transition to the
short-range universality class.

However, as we have seen in the case of the superconductor-insulator
transition, the effect of the metallic gate may well be more involved,
and it cannot automatically be assumed that its sole effect is to
screen the Coulomb interactions. Indeed, the general arguments
in Section \ref{npt} (relying as they did only on gauge invariance and the non-analytic 
form of the gauge field action) strongly suggest
that a similar fixed line may  also control the generic 
quantum Hall transition in the presence of a metallic plane.

To provide some calculational evidence for this suggestion,
we consider a simplified model used
 to describe the
quantum Hall transition with interacting particles, although the
effects of disorder, which are believed to be crucial for the physical
transition, are ignored. The model describing the critical point
consists of a massless Dirac fermion coupled to a Chern-Simons gauge
field \cite{mpafwu} and interacting via the Coulomb force
\cite{yesubir}.  This theory can be used to model both integer and
fractional Hall transitions by changing a statistics parameter.  The
Euclidean action for this model, in the absence of the metallic gate,
takes the form:
\bea
\nonumber
S &=& S_f + S_{gauge}\\\nonumber S_f &=& \int d^2xd\tau
\sum_{m=1}^N\psibar_m[\gamma_0(\partial_0-\frac{i}{\sqrt{N}}a_0)\\
& &
\;\;\;\;\;+\gamma_j(\partial_j-\frac{i\theta}{\sqrt{N}}a_j)]\psi_m\\\nonumber
S_{gauge} &=& \intk
[\tilde{a}_0(k,\omega)\epsilon_{ij}k_i\tilde{a}_j(-k,-\omega) \\ &
&\;\;\;\;+ \frac{e^{*2}}2
\tilde{V}_{Coul}(k)|\epsilon_{ij}k_i\tilde{a}_j(k,\omega)|^2]
\eea
where the $\gamma_\mu$ are the Dirac matrices in three dimensional
Euclidean spacetime (one representation in terms of the familiar Pauli
matrices is [$\gamma_0=i\sigma_3$; $\gamma_j=i\sigma_j$]), $e^*$ is
the Coulomb charge, $\theta/N$ the statistical angle and
$\tilde{V}_{Coul}(k)=1/k$ corresponds to a Coulomb potential of
$\frac1{2\pi r}$. In order to exert some control over the physics of
the critical point, $N$ species of fermions were introduced.  When
$N=1$ the model describes charged particles at the critical point
between an insulator and a quantized Hall state with Hall conductivity
$\sigma_{xy}=\frac{e^2}{h}(1-\frac{\theta}{2\pi})$. For $\theta=0$
this is a model of the integer transition, while $\theta=2\pi$
describes a Mott insulator to superfluid transition. In reference
\cite{yesubir} the model was considered in the limit of large $N$.
The results of those investigations to order $1/N$ were as
follows. The critical properties are controlled by a line of fixed
points parametrized by $\theta$ (the statistics is not renormalized,
as expected). For $\theta<1.24$, the Coulomb interaction is found to
be marginally irrelevant, while for $\theta>2$, it flows to strong
coupling. In between, $1.24<\theta<2$, the critical points are found
to have a finite Coulomb charge and $z=1$.

We now consider a system undergoing an insulator-quantized Hall state
transition in the presence of a metallic 2DEG. The geometry is taken
to be identical to figure \ref{fig1}, except that in place of the JJA,
a layer undergoing the quantum Hall transition is present. The
metallic 2DEG below it is assumed to have sufficiently high density of
electrons that we can ignore the effect of the applied magnetic field,
and treat it as a diffusive metal.

To model the quantum Hall transition, we use the anyon Mott transition
model described above, but now the particles in the quantum Hall layer
interact with the electrons in the metal, via the Coulomb force. Once
again we integrate out the diffusive electrons in the metal to obtain
an effective action for the quantum Hall transition. Again, only the
part of the response that is quadratic in the gauge fields is
retained, while higher order terms are neglected. It is convenient for
our purposes to write down this action in a somewhat different form
from that shown above.  An additional temporal gauge field $A_0$ is
introduced, and the Euclidean action can then be written as:
\bea
\nonumber
S &=& S_{Dirac} + S_{gauge}\\\nonumber S_{Dirac} &=& \int d^2xd\tau\,
\sum_{m=1}^N\psibar_m[\gamma_0(\partial_0-\frac{i}{\sqrt{N}}(a_0+A_0))\\
&&\;\;\;\;\;+\gamma_j(\partial_j-\frac{i\theta}{\sqrt{N}}a_j)]\psi_m\\\nonumber
S_{gauge} &=& \intk
[\tilde{a}_0(k,\omega)\epsilon_{ij}k_i\tilde{a}_j(-k,-\omega) \\ & &
\; \; \; \;+ \frac1{2e^{*2}}
\tilde{V}^{-1}_{eff}(k,\omega)|\tilde{A}_0(k,\omega)|^2]
\label{gauge}
\eea
where
\be
\tilde{V}^{-1}_{eff}(k,\omega) = |k| + \sigmatwo\frac{k^2}{|\omega|}
\ee
Note that gauge invariance and the non-analytic form of the terms
entering the gauge part of the action (\ref{gauge}) ensure that the
statistical angle $\theta$ as well as the term containing the
conductivity $\sigmatwo$ are not renormalized, and the Coulomb charge
flows according to the equation (\ref{flow}). Thus, there are two
strictly marginal operators parametrized by $\theta$ and
$\sigmatwo/e^{*2}$. We already know that the statistics parameter
generates a fixed line of critical points. We would now like to
address the question: under what circumstances, at a fixed value of
$\theta$, will the 2DEG also give rise to a fixed line? We will
address this question within the large-$N$ approximation. As we have
seen in earlier cases, if the Coulomb interaction turns out to be
marginally irrelevant, then there is a single fixed point controlling
the transition (at a fixed $\theta$) despite the fact that the term
containing the 2DEG conductivity $\sigmatwo$ is a strictly marginal
operator. This is because the gauge field $A_0$ decouples from the
problem in this limit. We would therefore like to calculate the flow
of the Coulomb charge to O($1/N$), in the presence of the metal. This
can be done following reference
\cite{yesubir}, but with the modified action (\ref{gauge}). For example, 
for the integer transition $\theta=0$, the flow equation for the
Coulomb charge in the presence of the metal can be easily worked out
($w=e^{*2}/16$):
\be
\frac{dw}{d\ln b} = -\frac{8w^2}{\pi^2}\int_0^{\pi/2}d\theta 
\frac{\sin^2\theta}{1+w\sin^2\theta+\sigmatwo\tan\theta}
\ee
the right-hand side clearly is always negative and hence the Coulomb
interaction at this integer transition is marginally irrelevant even
in the presence of the metal.

The flow equations for general $\theta$ and $\sigmatwo$ are
complicated and we do not attempt to write them out here. Instead we
will argue that at least for small enough $\sigmatwo$ there is a range
of $\theta$ fixed points for which, if we fix $\theta$, a fixed line
parametrized by $\sigmatwo$ can be obtained. First consider the case
of $\sigmatwo=0$ where a finite value of the Coulomb charge was found
for fixed points in a certain range: $1.24<\theta<2$
\cite{yesubir}. Imagine fixing a particular value of $\theta$ inside
this range. Then, for sufficiently small $\sigmatwo>0$, a fixed point
with finite Coulomb charge is still to be expected.  The strictly
marginal operator contained in the $\sigmatwo$ term will affect the
universal properties of the fixed point by inducing a fixed line at
this particular value of $\theta$. Since the renormalized Coulomb
charge is expected to remain finite along this fixed line, $z=1$ will
obtain. In addition to these fixed points, we can consider the effect
of the metal (with $\sigmatwo$ small) on the fixed points for
$\theta>2$. There, the $e^*\rightarrow \infty$ behaviour in the model
without the metallic plane, is expected, in the presence of a metallic
plane, to lead to a line of $z>1$ fixed points. These fixed points
should also be accessible within the large N technique; although the
Coulomb charge flows to infinity the gauge field fluctuations are
small and controlled by the marginal coupling $e^{*2}/\sigmatwo N$.

Thus, in this model of quantum Hall transitions, as in the
superconductor-insulator transitions discussed previously, the
proximity of the 2DEG can give rise to a line of fixed points
controlling the critical properties and hence affect the phase
transition in a non-trivial fashion.
 
\section{Comments on local dissipation models}
In
this section, we briefly comment on  previous theoretical
studies
 of Josephson arrays with ``local
dissipative baths'' whose results bear superficial similarities with those in the present paper. 
It is important to first realize that
we have focused on a physical situation where the origin of the dissipation 
is very clear. It is due to the coupling between the Cooper pairs and the gapless diffusive electrons in the metal. 
In contrast, Ref. \cite{Wgblst2} postulated the presence of dissipation due to some unspecified local degrees of freedom.
That paper also
suggested a fixed line controlling the
 transition
between the superconducting and insulating phases, once the
dissipation exceeds a critical strength $\alpha_0 = 2/3$.  The
dissipative term for each superconducting grain is assumed to be

\begin{equation} 
S_d = \alpha \int\,d\tau\,d\tau^\prime { (\phi(\tau) -
\phi(\tau^\prime))^2\over |\tau-\tau^\prime|^{2}}. \label{singlejunct}
\end{equation}

This model is not expected to describe the experiments in Refs.
~\cite{Rimberg,Mason}.  The zero diffusion constant limit of our model
resembles superficially the local dissipation models considered in~\cite{Kivelson}, 
which in turn differs from (\ref{singlejunct}) in
that the local dissipation is coupled to phase differences across
junctions.  Moreover, in our case the crucial 
boundary condition on the phase in the imaginary time direction 
(\ref{effaction}) can differ from that derived for more phenomenological 
models of local dissipation~\cite{Kivelson,Fisherdiss}. Furthermore,
 the presence
of long ranged interactions is crucial to the physics discussed in this
paper.

Finally we point out a problem with the approximation used in~\cite{Wgblst2} 
to derive the presence of a fixed line in the absence of both long-ranged 
interactions and disorder.  We show in Appendix \ref{appendix0}, that when exactly
the same approximation is applied to the 3D anisotropic XY model, one 
erraneously concludes the existence of a fixed line in that problem, 
in which the critical properties are actually controlled by a 
single fixed point.

\section{Implications for Experiments}
\label{exp_imp}
In this section we will consider the physical consequences of the
field theoretical results derived earlier. We first address a
quantitative question - at what value of the 2DEG conductivity are the
critical properties of the superconductor-insulator transition
significantly affected? For the $z=1$ fixed points, this will clearly
occur when the value of the marginal coupling ($1/g$) is of the same
order or larger than the renormalized Coulomb coupling
($1/e^{*2}_c$). Reinstating the physical parameters, this yields the
following condition on the 2DEG conductivity in SI units:
$\sigmatwo^{phys} > (\frac{e^{*2}}{e_c^{*2}})2c_s\kappa\epsilon_0$,
where $\kappa\epsilon_0$ is the permittivity constant and $c_s$ is the
bare speed of sound in the superfluid (in the absence of Coulomb
interactions).  If we make the naive assumption that the renormalized
Coulomb force is roughly the same as the bare Coulomb force
$e^*_c\approx e^*$, then we can obtain an estimate for the 2DEG
conductivity at which a significant effect on the critical properties
is felt. The speed of sound is given by ($c_s\sim
\frac{a}{\hbar}\sqrt{E_J E_C}$), where $a$ is the distance between
superconducting grains and $E_J$ and $E_C$ are the Josephson coupling
energy and the charging energy of a grain. Typically, at the
transition $E_J\sim E_C$ (if the 2DEG conductivity is not too large)
and so we obtain $c_s\sim E_Ca/\hbar$. Finally, substituting the form
of the charging energy for a grain of radius $R$,
$E_C=e^{*2}/4\pi\kappa\epsilon_0R$, we obtain that the nondimensional
conductivity should be $2c\kappa\epsilon_0
\sim \frac{e^{*2}}{h}(a/R)$. This is roughly of order the universal 
conductivity, if the grain size is approximately the same as the
distance between grains. Hence, we expect the effect of the 2DEG to be
substantial once its conductivity is of order
(6.4k$\Omega$)$^{-1}$. This regime is readily accessible in
experiments.


An alternate estimate of this conductivity for the $z=1$ fixed line
can be derived from the epsilon expansion solution to the model with
particle-hole symmetric disorder. There, we know that the exponents of
the $z=1$ fixed points are strongly affected once the marginal
coupling is of the order of $1/g_c=1.7\times 10^{-3}\epsilon$. Setting
$\epsilon=1$ and using the relation $\sigmatwo^{phys} =
(2\pi)\frac{e^{*2}}{h} \frac1g$, we obtain $\sigmatwo^{phys} = 0.01
\frac{e^{*2}}{h} \approx (600k\Omega)^{-1}$. This corresponds to a
smaller value of the conductivity than the earlier estimate, because
the Coulomb charge at the fixed point is strongly renormalized. For
the $z>1$ fixed points, the marginal coupling controls the gauge field
fluctuations. However, once the 2DEG conductivity greatly exceeds the
quantum of counductance $1/g \gg 1
\Rightarrow \sigmatwo \gg (2\pi)\frac{e^{*2}}{h} = (1.1k\Omega)^{-1}$ the 
critical exponents approach those of the short-ranged model without a
metallic plane.

 We now briefly consider how the results for the fixed line of
critical points obtained in this paper may be tested experimentally,
both for the superconductor-insulator transition as well as the
quantum Hall transition. Ideally, one would like to investigate the
universal critical properties at the transition, such as the critical
exponents and the conductivity at the transition, as a function of the
2DEG (metallic ground plane) conductivity. A fixed line of critical
points implies that these properties are a function of where the phase
boundary is crossed. Although these various `universal' quantities
would vary along the fixed line, there would be universal relations
amongst them, since they can all be expressed as a function of a
single parameter, the strength of the marginal coupling.  We also note
the caveat, that for large values of the 2DEG conductivity, there will
be region about the critical point where the system will behave like
the short-ranged interaction model, before ultimately crossing over to
the true fixed point behaviour very close to the transition.

Finally, we note that in a finite sized system (such as the JJA in
reference \cite{Rimberg} which is a 40x40 system) the transition is
rounded off when the correlation length exceeds the linear size of the
system. The width of this region ($\Delta g$), in terms of a control
parameter $g$ that tunes the transition, is given by the exponent
$\nu$ ($\Delta g\sim L^{-1/\nu}$, where $L$ is the linear system
size). The double epsilon expansion described earlier, on a model with
particle-hole symmetric disorder that exhibits the superconductor to
insulator transition, revealed a strong variation of $\nu$ along the
fixed line.  Such a variation could presumably be detected by studying
the width of the transition in these finite sized systems.

\section{Conclusions}

In this paper, we have studied the effects of a proximate metallic
plane on various two dimensional localization transitions with an
emphasis on the superconductor-insulator transition. Such a metallic
plane has two principal effects---it provides a mechanism for
screening the long-ranged Coulomb interaction, and it is a source of
dissipation due to the gapless diffusive electrons. The interplay of
these two effects leads to interesting physical phenomena. Perhaps the
most interesting result in this paper is the possibility of a fixed
line with variable critical exponents controlling the transition. In
addition, right at the transition, a temperature independent
conductivity that will vary along the fixed line is expected.

Our results are of direct relevance to experiments probing the
superconductor-insulator transition in Josephson junction arrays in
the presence of a metallic plane \cite{Rimberg}. So far these
experiments have not probed the universal scaling regime near the
transition. We hope that our results will focus future experimental
work on this regime.

A similar scenario may be expected to hold for quantum Hall
transitions in the presence of a proximate metallic plane. A line of
fixed points would again imply variable critical exponents and a
temperature independent value for the diagonal conductivity at the
transition. Recently, quantum Hall systems with nearby metallic planes
have been prepared \cite{screening} and it would be interesting to
experimentally investigate the effect of the metallic plane on the
critical properties at the quantum Hall transitions.

One limitation of our field theoretic approach is that, following
Hertz\cite{Hertz}, we have integrated out the gapless diffusive
electrons to obtain an effective action for the critical bosonic
Cooper pair degrees of freedom. As with other problems involving
quantum phase transitions in fermionic systems, a more satisfying
theoretical approach would keep both gapless fermionic and bosonic
modes as part of the effective theory and treat them on equal
footing. Unfortunately, such an approach is not available at present.

Another assumption that has been made which is needed to justify an
approximation used through this paper is that the fixed points satisfy
$z<2$. We leave it to future work to investigate whether
superconductor to insulator transitions in the presence of Coulomb
interactions and a metal can be controlled by fixed points with
dynamical exponents $z\ge2$, and if so what their description is.
 
We conclude by noting that our results show that even a simple Coulomb
interaction between the Cooper pairs and the gapless fermionic degrees
of freedom has a profound effect on the universality class of the
superconductor-insulator transition. Should the fixed line found in
this paper survive the inclusion of processes where the Cooper pair
can decay into two electrons, it will lead to a temperature
independent though apparently non-universal conductivity at the
superconductor-metal transition.

\section{Acknowledgements}
It is a pleasure to thank Leon Balents, Matthew Fisher, Akakii Melikidze, Andrew Millis
and Subir Sachdev for useful comments and criticism. A.V. would like
to thank Bell Laboratories for hospitality during an early stage of
this work, and the Pappalardo Fellows program at MIT for
support.  J. E. M. was supported by the LDRD program of Lawrence Berkeley National Laboratory.  T.S. was supported by the MRSEC program of the National
Science Foundation under grant number DMR-9808941, and by the NEC Corporation Fund. 

\appendix

\section{Large N Calculation of $z-1$ in the clean case.}
\label{appendixa}
In this section we describe in some detail how the dynamical critical
exponent $z$ can be calculated within the large-$N$ approximation, in
the absence of disorder.

We consider the problem directly in $D=2$, but with $N$ species of
bosons coupled to the scalar component of a gauge field as described
by equation (\ref{SlargeN}). The bare propagators for the gauge field
and the bosons are:
\bea
{\mathcal G}_A^{0}(k,\omega) &=& \frac{e^{*2}}{N} (|k| + \sigma_e
\frac{k^2}{|\omega|})^{-1}\\
{\mathcal G}^{0}(k,\omega) &=& (q^2 +\omega^2)^{-1}
\eea

\begin{figure}
\epsfxsize=2.8in
\centerline{\epsffile{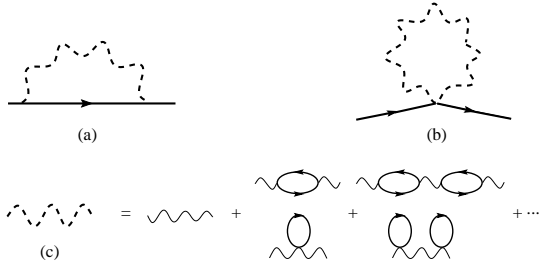}}
\vspace{0.1in}
\caption{To evaluate the O(1/N) correction to the dynamical critical 
exponent; only diagram (a) needs to be considered, where the solid
line is the boson propagator at criticality and the dashed-wavy line
is the RPA form of the gauge propagator, defined in (c), that appears
to this order in the large-$N$ approximation.}
\label{largeN}
\end{figure}

Since it is only the dynamics and coupling of the gauge field that
distinguishes space and time, any contribution to $z-1$ must come
entirely from this interaction. We have already noted in Section
\ref{section2D} that to leading order in 1/N (saddle point) the gauge
field decouples from the problem, and the dynamical critical exponent
stays equal to unity at this stage. Therefore, we calculate this
quantity to O(1/N). First, we recognize that the gauge field
propagator will be dressed by boson density fluctuations as shown in
Figure \ref{largeN}a, so that we obtain the RPA form:
\be
{\mathcal G}_A^{-1}(k,\omega) = [{\mathcal G}_A^{0}(k,\omega)]^{-1} +
\Pi^{0}(k,\omega) 
\nonumber
\ee
where $\Pi^0$ is the contribution from the pair of boson bubble
diagrams in Figure \ref{largeN}c
\bea
\nonumber
\Pi^{0}(k,\omega) &=& \intK {\mathcal G}_0(K,\Omega) \\
&& [(\omega+2\Omega)^2{\mathcal G}_0(k+K,\omega+\Omega) - 1]\\ &=&
\frac1{16}\frac{k^2}{\sqrt{k^2+\omega^2}}
\eea
In order to calculate $(z-1)$ to O(1/N), we need to evaluate the self
energy contribution arising from the diagrams in Figure
\ref{largeN}b. We will be interested in isolating terms in the self
energy that are of the form:
\be
\Sigma(k,\omega)\sim \eta k^2\ln(\sqrt{k^2+\omega^2}/\Lambda)+\eta_\omega 
\omega^2 \ln(\sqrt{k^2+\omega^2}/\Lambda)
\label{selfenergy}
\ee
where $\Lambda$ is the ultraviolet cutoff. These contributions will
modify the boson propagator since:
\be
{\mathcal G}^{-1}(k,\omega) = [{\mathcal G}^{0}(k,\omega)]^{-1} -
\Sigma(k,\omega)
\ee
and thus affect the power laws that characterise the boson correlators
at criticality. The dynamical critical exponent is then given by:
\be
z-1 = \frac12(\eta_\omega -\eta)
\ee 
Notice, that since we are only interested in contributions of the form
(\ref{selfenergy}), which vanish in the zero frequency limit, we only
need to calculate $\Sigma(k,\omega)-\Sigma(0,0)$, therefore the second
diagram in Figure \ref{largeN}b, which gives a momentum independent
contribution, can be ignored.  Evaluating the resulting diagram:
\bea
\nonumber
&&\Sigma(k,\omega)-\Sigma(0,0) = \\\nonumber
\frac1N\intK [&&(2\omega+\Omega)^2{\mathcal G}^{0}(k+K,\omega+\Omega)\\
&&-\Omega^2{\mathcal G}^{0}(K,\Omega)]{\mathcal G}_A(K,\Omega)
\eea
we can obtain the correction to the dynamical critical exponent to
O(1/N), whic is:
\begin{eqnarray*}
z-1 &=& \frac1N I(e^{*2},\sigmatwo)\\ I(e^{*2},\sigmatwo)&=&
\int_0^{\pi/2} \frac{d\theta}{2\pi^2}
\frac{\sin^2\theta(1-3\sin^2\theta)}{e^{*-2}(1+\sigmatwo\tan\theta)+\frac{\sin\theta}{16}}
\end{eqnarray*}
Note, that this can be shown to imply that $z-1>0$ for all $e^*>0$,
$\sigmatwo>0$ and hence the only fixed point is at $e^*=0$.

\section{The $\epsilon$, $\epsilon_\tau$ Expansion}
\label{appendixb}
We consider the problem in $D=3-\epsilon$ spatial dimensions, with
particle-hole symmetric disorder correlated in $\epsilon_\tau$
dimensions, given by the action in equation (\ref{sdis}). We consider
integrating out all modes with wavevectors in the range
$[\Lambda/b,\Lambda]$. The additional terms generated lead to a
renormalization of the terms in the action; the quadratic derivative
terms are now multiplied by the scale factors $Z_\omega$, $Z_k$ and
the other terms acquire factors $Z_r$, $Z_u$ and $Z_{\bar{W}}$. Note,
in dimensions $D<3$, the Coulomb term, as well as the term arising
from coupling to the 2DEG in the gauge propagator are unchanged on
integrating out high frequency modes since they are non-analytic in
frequency wave-vector space. On rescaling the momenta $k'=bk$ to
restore the cutoff to $\Lambda$ and also rescaling $\omega'=b^z\omega$
and the fields $\psi'=Z_k^{\frac12}\psi$ and
$\tilde{A}'_0=b^{-D}\tilde{A}_0$ (the gauge field scaling follows from
gauge invariance) to obtain a new set of couplings at this scale:
\begin{eqnarray}
\label{scalesu} 
u' &=& b^{\epsilon+1-z}Z_k^{-2}Z_u u\\
\label{scalesW}
\bar{W}' &=& b^{\epsilon + 1-z+\epsilon_\tau}Z_k^{-2}Z_{\bar{W}}\bar{W}\\
e^{*'2} &=& b^{z-1}e^{*2}\\
\label{scalesr}
r' &=& b^2 Z_k^{-1} Z_r r
\end{eqnarray}
In the following, we will look for fixed point solutions to the above
flow equations, assuming that $\epsilon$, $\epsilon_\tau$ are
small. In that limit, we are justified in looking for fixed points
where the couplings are small, and hence a perturbative evaluation of
the flows is feasible.

\subsection{Calculation of $z$ and $\eta$; Flow of $e^*$.}
\label{sectionselfenergy}
Here we calculate the change in the quadratic derivative terms ($Z_k$,
$Z_\omega$) on integrating out the high momentum modes. In practice,
this is most easily done by integrating all modes below a cutoff
$\Lambda$, and extracting the log divergent contributions to the boson
propagator ${\mathcal G}$:
\bea
\nonumber
{\mathcal G}^{-1}_{ren}(k,\omega) = k^2 + \omega^2 &-& \eta k^2
\ln(\sqrt{k^2+\omega^2}/\Lambda)\\ 
&-& \eta_\omega \omega^2 \ln(\sqrt{k^2+\omega^2}/\Lambda)
\label{term}
\eea
More details on how these two proceedures are related to each other
may be found in \cite{Shankar}. Here, we just note that if integrating
all modes below the cutoff $\Lambda$ gives rise to the additional term
in equation (\ref{term}), then the contribution on just integrating
modes in a shell betwen $[\Lambda/b,\Lambda]$, can be found by
subtracting the same term but with $\Lambda\rightarrow \Lambda/b$
which yields:
$$
(\eta k^2 +\eta_\omega \omega^2)\ln b
$$ 
. These can then be related to the quantities we want to calculate:
\bea
\label{Zk}
Z_k &=& 1+ \eta \ln b\\ Z_\omega &=& 1 + \eta_\omega \ln b
\label{wavefn}
\eea

\begin{figure}
\epsfxsize=2.8in
\centerline{\epsffile{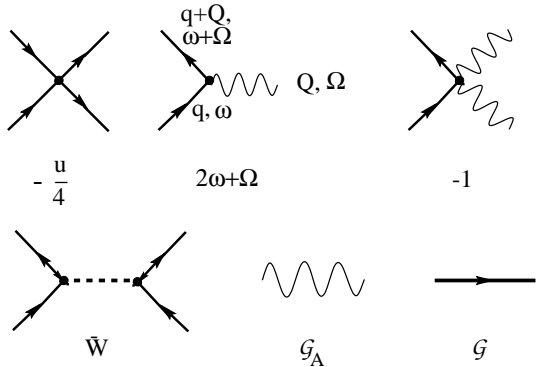}}
\vspace{0.1in}
\caption{Feynman rules used in the $\epsilon$, $\epsilon_\tau$ expansion. 
The vertices arising from the $|\psi|^4$ interaction, the gauge
interaction and disorder generated interaction are shown. The dashed
disorder line only transfers the $D+1-\epsilon_\tau$ spatial
components of the momentum.}
\label{rules}
\end{figure}

The Feynman rules derived from the action (\ref{sdis}) are shown in
Figure
\ref{rules}, where the boson propagator at criticality is given by 
${\mathcal G}^{-1}=(k^2+\omega^2)$ and the gauge propagator is given
by ${\mathcal G}_A^{-1}
=e^{*-2}(k^{D-1}+\sigmatwo\frac{k^D}{|\omega|})$.

\begin{figure}
\epsfxsize=2.8in
\centerline{\epsffile{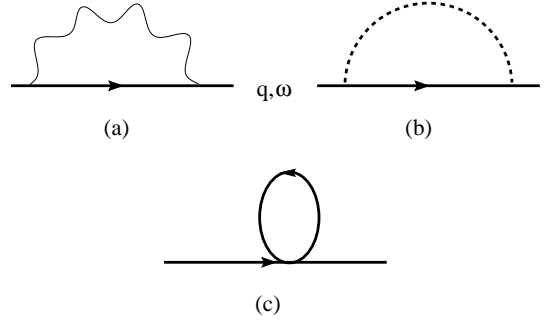}}
\vspace{0.1in}
\caption{The diagrams (a) and (b) contribute to renormalizing the the 
quadratic derivative terms in the action, while all three diagrams
contribute to the renormalization of the quadratic term $r$.}
\label{figselfenergy}
\end{figure}

We are now in a position to calculate the quantities in
(\ref{wavefn}).  Since we are interested in the momentum dependent
contribution to the self energy, we only need to calculate
$\Sigma(k,\omega)-\Sigma(0,0)$. The two diagrams of interest then are
shown in Figure \ref{figselfenergy}a and \ref{figselfenergy}b. We work
to lowest order in $\epsilon$ and therefore the integrals will be
performed in $D=3$, and the terms that depend on the logarithm of the
cutoff will be extracted. The contribution from gauge fluctations
(Figure
\ref{figselfenergy}a) is:
\bea
\nonumber
\Sigma(k,\omega)-\Sigma(0,0) &=& \intKthree {\mathcal G}_A(K,\Omega)\\
& &[(2\omega+\Omega)^2{\mathcal
G}(k+K,\omega+\Omega)-\Omega^2{\mathcal G}(K,\Omega)]
\eea 

which yields:
\bea
\eta^{A_0}_\omega &=& 
-\frac{e^{*2}}{2\pi^2}\int_0^{\frac\pi2}\frac{d\theta}{\pi}
\frac{4-9\cos^2\theta+4\cos^4\theta}{1+\sigmatwo \tan\theta}\\
\eta^{A_0} &=& \frac{e^{*2}}{6\pi^2}\int_0^{\frac\pi2}\frac{d\theta}{\pi} 
\frac{(4\cos^2\theta-1)\cos^2\theta}{1+\sigmatwo \tan\theta}
\eea

Similarly, the contribution arising from the disorder diagram in
Figure
\ref{figselfenergy}b can be calculated.  Since the disorder induced 
interaction is independent of spatial momentum transfer,
$\Sigma(q,0)=\Sigma(0,0)$, the disorder contribution to $\eta$
vanishes.  However, at finite frequency transfer we have:
\be
\Sigma(0,\omega)-\Sigma(0,0) = -\omega^2\bar{W}\int 
\frac{\dcross^{D+1-\epsilon_\tau}Q}{Q^4}
\ee
which, to lowest order in $\epsilon_\tau$ can be evaluated to give
\be
\eta_\omega^{\bar{W}} = - \frac{\bar{W}}{8\pi^2}
\ee
putting this all together we have
\begin{eqnarray*}
\eta &=&\eta^{A_0}\\
\eta_\omega&=&\eta_\omega^{A_0} + \eta_\omega^{\bar{W}}
\end{eqnarray*}
from this the flow equation for the Coulomb charge is easily derived:
\bea
\beta(e^*)=\frac{de^{*2}}{d\ln b} &=& (z-1)e^{*2}\\
z-1 &=& \frac12(\eta_\omega - \eta)
\label{zminusone}
\eea

after redifining the couplings $q^2=\frac{e^{*2}}{2\pi^2}$ and
$W=\frac{\bar{W}}{2\pi^2}$ we get equation (\ref{coulomb}).

\subsection{Flow Equations for $u$:}
\begin{figure}
\epsfxsize=2.8in
\centerline{\epsffile{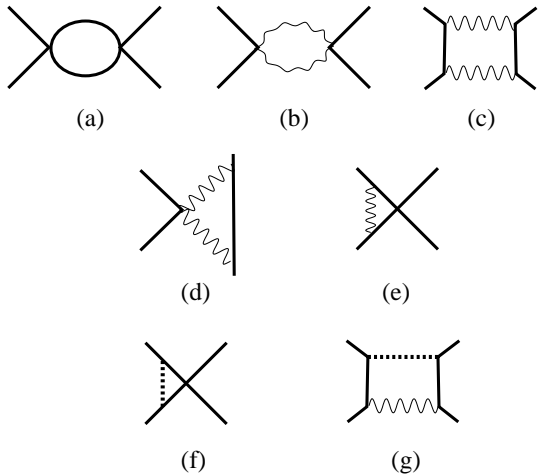}}
\vspace{0.1in}
\caption{The diagrams contributing to the renormalization of the quartic 
interaction.}
\label{renormu}
\end{figure}

We begin by calculating the renormalization of the quartic term ($u$)
arising from the integration of high frequency modes
($Z_uu$). Diagrams Figure \ref{renormu}a-g contribute, and we can
write the result as $Z_uu=u+$(Fig. \ref{renormu}a+..+g)$\ln b$. We
will evaluate these diagrams and extract the log divergent parts; the
external lines are assumed to carry zero momentum.

	The contribution from the first diagram can be found in any
standard reference on the $\epsilon$ expansion for short-ranged model
\cite{Shankar}, and is given by:
$$
\mbox{\ref{renormu}a} \rightarrow -\frac5{2} \frac{u^2}{8\pi^2}
$$
 Next, contributions arising from the gauge interaction (Figure
\ref{renormu}b-e) can be evaluated. Details about the combinatorial 
factors that appear for these diagrams are not repeated here as they
as they are discussed in \cite{Ye}; the only difference for our
calculations is that the gauge propagator is modified due to coupling
to the 2DEG. The result for these diagrams then is
$$
\mbox{\ref{renormu}b+\ref{renormu}c+\ref{renormu}d} \rightarrow 
-\frac{e^{*4}}{2\pi^2}[\frac2\pi\int^{\pi/2}_{0}d\theta\frac{\sin^2\theta}{(1+\sigmatwo\tan\theta)^2}]
$$
while the graph in Figure \ref{renormu}e contributes:
$$
\mbox{\ref{renormu}e} \rightarrow 
\frac{ue^{*2}}{4\pi^2}[\frac4\pi\int^{\pi/2}_{0}d\theta 
\frac{\cos^2\theta}{1+\sigmatwo\tan\theta}]
$$  
The diagram in Figure \ref{renormu}f which involves the disorder can
be computed in the following way. There are $^4\!C_2$ diagrams of the
type shown, and hence the contribution is:
$$
\mbox{\ref{renormu}f} =^4\!\!C_2 u\bar{W}\int \frac{\dcross^4p}{p^4} 
\rightarrow\frac{3u\bar{W}}{4\pi^2}
$$
Finally, the diagram in Figure \ref{renormu}g makes no
contribution. Since the external lines are at zero frequency, and the
disorder interacton, being independent of time, does not change the
frequency, the coupling to the gauge field vanishes.

Putting together these results, as well as the factors arising from
rescaling the fields (\ref{scalesu}, \ref{Zk}, \ref{zminusone}) and
redefining the couplings as in (\ref{couplings}), we obtain the flow
equation (\ref{lambda}).

\subsection{Flow Equations for Disorder Coupling}
\begin{figure}
\epsfxsize=2.8in
\centerline{\epsffile{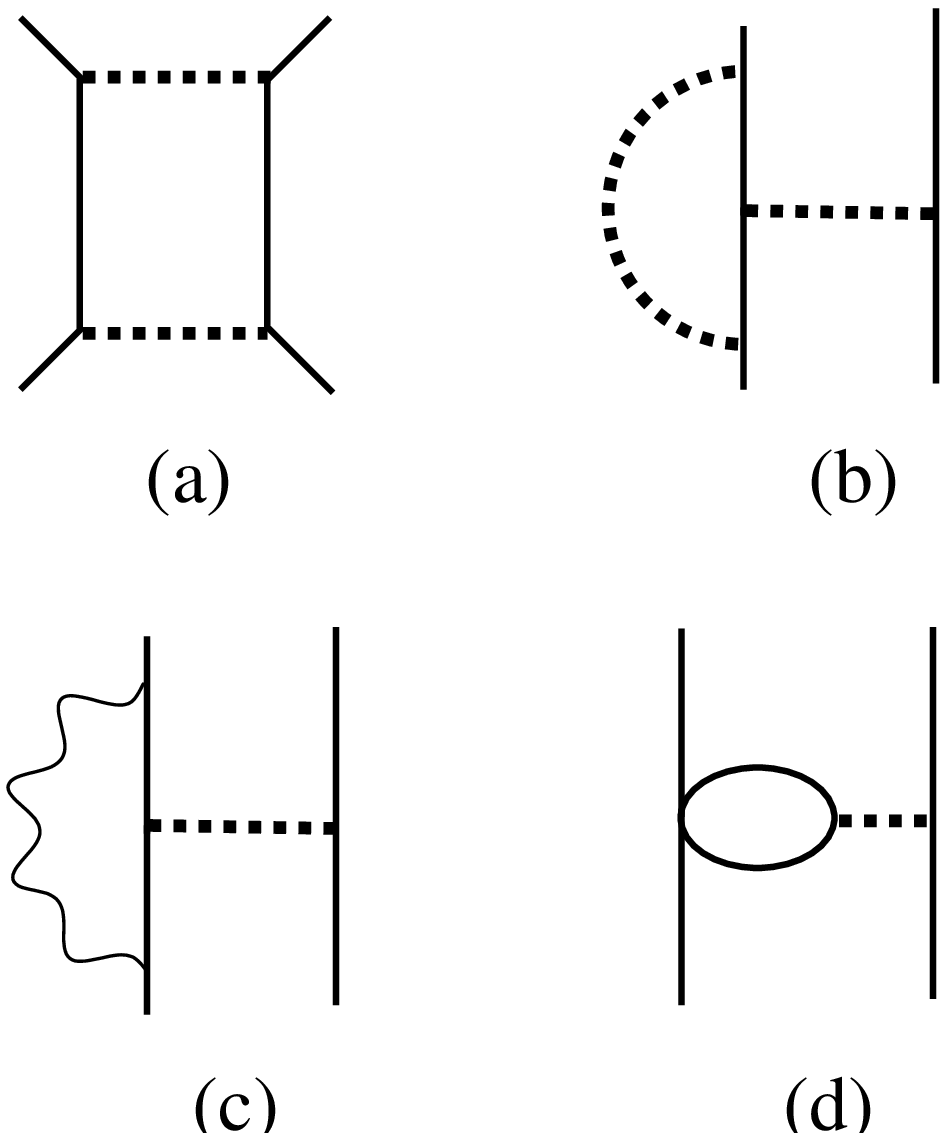}}
\vspace{0.1in}
\caption{The diagrams contributing to the renormalization of the 
interaction induced by disorder.}
\label{renormW}
\end{figure}

We first consider the renormalization of the disorder term ($\bar{W}$)
arising from the integration of high frequency modes
($Z_{\bar{W}}\bar{W}$). Diagrams Figure \ref{renormW}a-d contribute,
and we can write the result as $Z_{\bar{W}}\bar{W}=\bar{W}+$(Fig.
\ref{renormW}a+..+d)$\ln b$.

To evaluate the diagram in Figure \ref{renormW}a we note there are two
identical processes that contribute, with the boson lines parallel or
anti-parallel. Similarly, there are two equivalent processes
associated with the diagram in Figure \ref{renormW}b, with the
disorder induced interaction decorating either the left or right boson
line. These diagrams all make the same contribution; putting them
together:
$$
\mbox{\ref{renormW}a+\ref{renormW}b} = 4\bar{W}^2\int 
\frac{\dcross^4p}{p^4} \rightarrow \frac{\bar{W}^2}{2\pi^2}
$$ 
Similarly, for the diagram in Figure \ref{renormW}c there are two
equivalent contributions, since the gauge interaction can decorate
either the left or the right boson line. Hence,
\begin{eqnarray*}
\mbox{\ref{renormW}c}&=&2\bar{W}\intKthree \Omega^2 {\mathcal 
G}_A(K,\Omega){\mathcal G}^2(K,\Omega)\\ &\rightarrow&
\frac{e^{*2}\bar{W}}{4\pi^2}[\frac4\pi\int_0^{\pi/2}d\theta\frac{\cos^2\theta}{1+\sigmatwo\tan\theta}]
\end{eqnarray*}
Finally, the the diagram in Figure \ref{renormW}d also has two
equivalent realizations, and the boson self interaction loop can be
inserted in four ways, which yields:
$$
\mbox{\ref{renormW}d} = 8\bar{W}(-u/4)\int \frac{\dcross^4p}{p^4} 
\rightarrow -\bar{W}\frac{u}{4\pi^2}
$$
Putting together these results, as well as the factors arising from
rescaling the fields (\ref{scalesW}, \ref{Zk}, \ref{zminusone}) and
redefining the couplings as in (\ref{couplings}), we obtain the flow
equation for the disorder coupling (\ref{W}). This completes the
derivation of the flow equations (\ref{coulomb} - \ref{W}).

\subsection{The $z>1$ Fixed Line}
\label{appendixd}
The fixed points in the presence of a metallic plane in the case
$e^*\rightarrow \infty$ are obtained from the equations
(\ref{lambdaeinfinity}, \ref{Weinfinity}) assuming that the marginal
coupling $\bar{g}$ is $O(\epsilon$). This yields the fixed point
values of the couplings, (in terms of
$x(\bar{g})=\frac13(\frac4\pi\bar{g}+4\epsilon+11\epsilon_\tau )$):
\bea
\lambda_c &=& \frac13(x(\bar{g})+ \sqrt{\frac72\bar{g}^2+x^2(\bar{g})})\\
W_c &=& \frac87(2\lambda_c -\frac{\bar{g}}{\pi} - \epsilon
-\epsilon_\tau)
\eea
the couplings are $O(\epsilon)$ and the dynamical critical exponent is
given by:
\be
z=1+\frac{W_c}8 - \frac{\bar{g}}{3\pi}
\ee
These solutions exist for all $O(\epsilon)$ values of $\bar{g}$, and
have $z>1$, which ensures that $e^*=\infty$ does not flow. In the
linit $\bar{g}\rightarrow 0$ (which corresponds to a large value of
the 2DEG conductivity) we recover the short-range interacting boson
case, in the absence of a metallic plane (studied in
\cite{Cardy}). Note, the flow equations (\ref{lambdaeinfinity},
\ref{Weinfinity} with $\bar{g}=0$) however differ from those in
\cite{Cardy}. This difference is a result of treating the time
direction as distinct in the presence of Coulomb interactions
\cite{Herbut}, which leads us to scale the measure $[d\omega dk^D]$ as
$L^{-z-D}$ rather than as $L^{-\epsilon_\tau(z-1)-D}$ as in
\cite{Cardy} where only the $\epsilon_\tau$ dimensions are assumed to
scale with the dynamical exponent. Therefore when we continue the
fixed point obtained in the presence of Coulomb interactions to the
$\bar{g}=0$ fixed point, it differs in details from the result
obtained in \cite{Cardy} but agrees with the short-ranged fixed point
quoted in \cite{Herbut}.

\subsection{The Critical Exponent $\nu$} 
To calculate the exponent $\nu$, we need the flow equations for the
coefficient of the quadratic term ($r$). In fact, we only need those
terms in the flow equations that themselves contain $r$. The diagrams
that then need to be considered are those in Figure
\ref{figselfenergy}a,b and c, evaluated with zero momentum and
frequency in the external lines. In these, the terms proportional to
$r$ that contribute to its renormalization are:
\begin{eqnarray*}
\mbox{\ref{figselfenergy}a} &=& r\intKthree {\mathcal 
G}^2(K,\Omega)\Omega^2{\mathcal G}_A(K,\Omega) \\ & &\rightarrow
r\frac{e^{*2}}{2\pi^2}[\frac4\pi
\int_0^{\frac\pi2}d\theta\frac{\cos^2\theta}{1+\sigmatwo\tan\theta}]\\
\mbox{\ref{figselfenergy}b} &=& r\bar{W}\int \dcross^4p{\mathcal G}^2(p)\\
& &\rightarrow r\frac{\bar{W}}{8\pi^2}
\end{eqnarray*}
finally, evaluating the diagram in Figure \ref{figselfenergy}c, we
note that there are four ways to pick boson lines to make a loop and
hence:
\begin{eqnarray*}
\mbox{\ref{figselfenergy}c} &=& 4r(-\frac{u}4)\int\dcross^4p{\mathcal 
G}^2(p) \\ &\rightarrow& -r\frac{u}{8\pi^2}
\end{eqnarray*}
Combining this with the effect of rescaling the fields, as given in
equation (\ref{scalesr}) and rescaling the couplings according to
(\ref{couplings}), we obtain the following flow equation for this
relevant coupling.
\be
\frac{dr}{d\ln b} = [2-\lambda+\frac W4+\frac D{12} q^2]r + \dots 
\mbox{(terms independent of r)}
\ee
Where the coefficient $D$ is a function of $\sigmatwo$ and is defined
in equation (\ref{D}). At the fixed point this leads to the following
equation for the critical exponent $\nu$:
\be
\nu^{-1} = 2-\lambda_c +\frac{W_c}{4} +\frac{D}{12}q_c^2
\ee
since these fixed point couplings are all small [O($\epsilon$)], we
can invert this expression to obtain equation (\ref{nu}). In the limit
$\sigmatwo=0$, we can compare this answer with earlier work. The
coefficient of the $q^2_c$ term differs from the expression in
\cite{Herbut}, but agrees with that in \cite{Ye}

{\it The Exponent $\nu$ for the $z>1$ Fixed Line}

For this fixed line, $\nu$ can be derived from equation (\ref{nu}) by
formally letting $q^2, \sigmatwo \rightarrow\infty$ but holding the
ratio $q^2/\sigmatwo=\bar{g}$ finite. This yields:
\be
\nu = \frac12 + \frac14(\lambda_c(\bar{g}) -\frac{\bar{g}}{3\pi} 
-\frac{W_c}4(\bar{g}))
\ee 

\subsection{Stability of the Fixed Points:}
We would like to make sure that besides the relevant variable $r$, the
fixed point is stable to perturbing away in any of the other
couplings. A linear stability analysis can be performed on the fixed
point at $(q_c^2,\: W_c,\: \lambda_c)$ by move slightly away
$(q_c^2+\delta q^2,\: W_c+\delta W,\: \lambda_c + \delta \lambda)$ and
asking how the deviations evolve under resclaing. This leads to the
equation,
\be
\frac{d}{d\ln b} \left ( \matrix{\delta q^2 \cr \delta W \cr 
\delta\lambda} \right)
= {\bf M_c} \left ( \matrix{\delta q^2 \cr \delta W \cr \delta\lambda}
\right)
\ee
where
\be
{\bf M_c} = \left[ \matrix{\partial_{q^2}\beta(q^2) &
\partial_{W}\beta(q^2) & \partial_{\lambda}\beta(q^2) \cr 
\partial_{q^2}\beta(W) &\partial_{W}\beta(W) 
&\partial_{\lambda}\beta(W)\cr \partial_{q^2}\beta(\lambda)
&\partial_{W}\beta(\lambda) &\partial_{\lambda}\beta(\lambda)} \right]
\ee
where the derivatives of the beta functions are evaluated at the fixed
point. If the eigenvalues of ${\bf M_c}$ are all negative, then the
perturbation will die out on running the RG, and the fixed point is
stable.

{\it Stability of $z=1$ Fixed Points}

We have numerically evaluated these eigenvalues along the $z=1$ fixed
line parametrized by $0<\sigmatwo<1.99$, and find that all these
critical points are stable (we have assumed $\epsilon=\epsilon_\tau$
which is consistent with the limit of $\epsilon=1$, $\epsilon_\tau=1$
that we are finally interested in). A pair of the eigenvalues, it
turns out, is complex (and conjugate to each other), but they have
negative real parts.  The least negative eigenvalue ranges from $\sim
-0.5\epsilon$ (at $\sigmatwo=0$) to $\sim -0.3\epsilon$ (near
$\sigmatwo=1.99$).

{\it Stability of $z>1$ Fixed Points} Here, the stability matrix we
need to consider is just a $2 \times 2$ matrix since the condition
$z>1$ which is satisfied by these fixed points ensures $1/e^*=0$
remains stable. The eigenvalues of this matrix were evaluated
numerically (assuming $\epsilon=\epsilon_\tau$) for the full range of
$g$ (assumed $O(\epsilon)$) and were found to be complex but with
negative real parts. Thus, all these points on the $z>1$ fixed line
are found to be stable.

\section{The anisotropic XY model}
\label{appendix0}
This appendix shows that the arguments in~\cite{Wgblst2} for
a fixed line for the dissipation term
\begin{equation}
S_d = \alpha \int\,d\tau\,d\tau^\prime { (\phi(\tau) -
\phi(\tau^\prime))^2\over |\tau-\tau^\prime|^{2}},
\end{equation}
which unlike in our model do not depend on either
long-ranged interactions or disorder, predict incorrect behavior in a
simpler case and hence are unreliable compared to controlled
approximations such as the $\epsilon$-expansion.

The approach of~\cite{Wgblst2} predicts a continuous set of
transitions for a simple related problem, an anisotropic XY model,
where only one superconducting transition is present.  This suggests
that phase-squared dissipation likely introduces at most one new
universality class for the superconducting transition.  The subtle
problem in the derivation of the effective action in~\cite{Wgblst2} is
that only the leading (two-spin) term is kept in the
Hubbard-Stratonovich transformation used to decouple the junctions.

Consider a classical 3D system made of layered 2D XY models with
interplane coupling $J_z \cos(\theta_i - \theta_{i+1})$, where $i$ is
a layer index.  The action after taking the continuum limit in the
planes is \begin{equation} S_{\rm XY} = \sum_i \int\,d^2x\,[J_{xy}
(\nabla
\theta_i)^2 - J_z \cos(\theta_i-\theta_{i+1})]. \end{equation}

Here the couplings are defined to include $\beta = 1/kT$. If $J_z =
0$, then the system can have algebraic in-plane order for $J_{xy}$
above the Kosterlitz-Thouless transition.  In this algebraically
ordered phase, the spin-spin correlation $\langle S_{\bf r_1} \cdot
S_{\bf r_2} \rangle
\propto |{\bf r_1}-{\bf r_2}|^{-\alpha}$, for some value $0\leq\alpha\leq
1/4$ depending on the in-plane coupling.  Introducing a complex boson
field $\psi \equiv S_x + i S_y$ to decouple the planes,
\begin{eqnarray} S&=& \sum_i \int\,d^2x\,[J_{xy} (\nabla \theta_i)^2 +
|\psi_i-e^{i
\theta_i}|^2 \nonumber \\ &&- J_z (\psi_i^* \psi_{i+1} +\psi_{i+1}^*
\psi_i)/2]. \end{eqnarray}

Now the approach of \cite{Wgblst2} is based on an effective action for
the full 3D model with nonzero interplane coupling $J_z$, using the
above 2D correlations. The quadratic part of this action reads in
cylindrical coordinates \begin{equation} S_{{\rm eff}} = \int
k\,dk\,d\phi\,dk_z\,({k_z}^2 + k^{\alpha-2}+ m^2)
|\psi(k,\phi,k_z)|^2.
\label{twodim} \end{equation} Now this quadratic part is used in
~\cite{Wgblst2} to calculate the critical conductivity as support for
the claim of a continous set of universality classes.

\begin{figure}
\epsfxsize=2.8in
\centerline{\epsffile{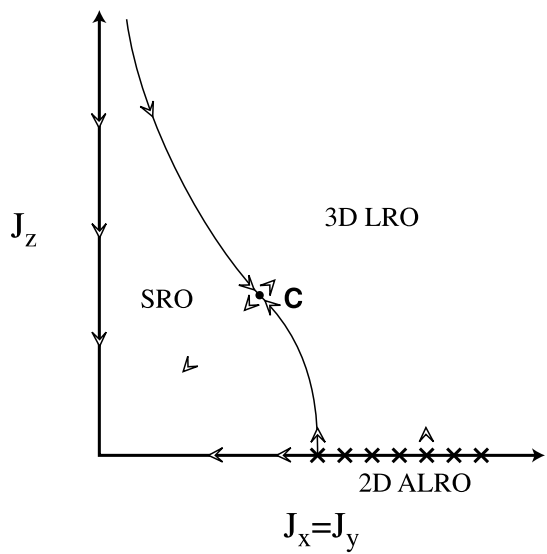}}
\vspace{0.1in}
\caption{Phase diagram of 3D XY model with direction-dependent couplings 
$J_z {\not =} J_x = J_y$.  Arrows indicate direction of projected RG flows.  
Despite the line of fixed points for $J_z=0$, all transitions between the 
long-range ordered (superconducting) phase and short-ranged (insulating) 
phase are controlled by the single critical point {\bf C}.
}
\label{figtwo}
\end{figure}

Even though (\ref{twodim}) reproduces the in-plane two-spin
correlation function in the absence of interplane coupling, it cannot
be used at this level of approximation to study the transition into
the superconductor.  The actual phase diagram of the anisotropic XY
model is shown in Fig.~\ref{figtwo}.  There is a single fixed point
that controls the superconducting transition for all nonzero values of
the couplings, with unique values of the critical exponents.

The above suggests that the model considered in~\cite{Wgblst2} may not
in fact have a continuous set of superconducting transitions once
$\alpha>\alpha_0= 2/3$.  The dissipation generates a series of of
terms in the $\psi$ action; all of these are irrelevant for small
$\alpha$, but the higher-order terms become relevant at the same time
as the leading term.  Weak phase-squared dissipation is indeed
irrelevant, but it is not clear what happens once the dissipation is
relevant.

\end{multicols}
\end{document}